\documentclass[%
reprint,
 groupedaddress,
superscriptaddress,
 amsmath,amssymb,
 aps,
 twocolumn,
]{revtex4-2}

\usepackage{mathrsfs}
\usepackage{textcomp}
\usepackage{units}
\usepackage{array,mathtools}
\usepackage{stmaryrd}
\usepackage{scrextend}
\usepackage{nicefrac}
\usepackage{multirow}
\usepackage{makecell}
\usepackage{bm}
\newcolumntype{C}{>{$}c<{$}}
\AtBeginDocument{
\heavyrulewidth=.08em
\lightrulewidth=.05em
\cmidrulewidth=.03em
\belowrulesep=.65ex
\belowbottomsep=0pt
\aboverulesep=.4ex
\cmidrulesep=\doublerulesep
\cmidrulekern=.5em
\defaultaddspace=.5em
}
\usepackage{booktabs}
\usepackage{ragged2e}
\usepackage{caption}
\usepackage{graphicx}% Include figure files
\usepackage{dcolumn}% Align table columns on decimal point

\usepackage{xr} % for external references if needed
\usepackage{hyperref}
\usepackage{cleveref}

\newcommand{\MainTitle}{Mechanisms of alkali ionic transport in amorphous oxyhalides solid state conductors}
\newcommand{\beginsupplement}{
    \setcounter{section}{0}
    \setcounter{table}{0}
    \setcounter{figure}{0}
    \renewcommand{\thesection}{S\arabic{section}}
    \renewcommand{\thetable}{S\arabic{table}}
    \renewcommand{\thefigure}{S\arabic{figure}}
}

\usepackage{xspace}

\usepackage[OT1]{fontenc}
\usepackage[cm]{sfmath}

\usepackage{multirow}
\usepackage{amsbsy}
\usepackage{bm}

\usepackage[english]{babel}

\usepackage{letltxmacro}
\LetLtxMacro{\ORIGselectlanguage}{\selectlanguage}
\makeatletter
\DeclareRobustCommand{\selectlanguage}[1]{%
  \@ifundefined{alias@\string#1}
    {\ORIGselectlanguage{#1}}
    {\begingroup\edef\x{\endgroup
       \noexpand\ORIGselectlanguage{\@nameuse{alias@#1}}}\x}%
}
\newcommand{\definelanguagealias}[2]{%
  \@namedef{alias@#1}{#2}%
}
\makeatother

\definelanguagealias{en}{english}
\definelanguagealias{EN}{english}
\definelanguagealias{american}{english} % add any others your .bib might use

\DeclareMathAlphabet{\mathsfbf}{OT1}{cmss}{bx}{n}

\newcommand{\s}{\sigma}

\usepackage[dvipsnames]{xcolor}
\usepackage[normalem]{ulem}

\newcommand{\editor}[2]{%
  \expandafter\newcommand\csname #1note\endcsname[1]{%
    \textcolor{#2}{(\textbf{#1:} ##1)}}%
  \expandafter\newcommand\csname #1\endcsname[1]{%
    \textcolor{#2}{##1}}%
  \expandafter\newcommand\csname #1cancel\endcsname[1]{%
    \textcolor{#2}{\sout{##1}}}%
  \expandafter\newcommand\csname #1change\endcsname[2]{%
    \textcolor{#2}{\sout{##1} ##2}}%
  \newenvironment{#1text}{\color{#2}}{\color{black}}
}

\editor{LB}{ForestGreen}

\begin{document}

\title{\MainTitle}
%\title{Microscopic theory of alkali diffusion in amorphous oxyhalide solid-state superionic conductors}

\author{Luca Binci}
\author{KyuJung Jun}
\author{Bowen Deng}
\author{Gerbrand Ceder}
\email{gceder@berkeley.edu}
\affiliation{Department of Materials Science and Engineering, University of California Berkeley, Berkeley, California, USA}
\affiliation{Materials Sciences Division, Lawrence Berkeley National Laboratory, Berkeley, California, USA}
\date{\today}

\begin{abstract}
Amorphous oxyhalides have attracted significant attention due to their relatively high ionic conductivity ($>$1 mS cm$^{-1}$), excellent chemical stability, mechanical softness, and facile synthesis routes via standard solid-state reactions. These materials exhibit an ionic conductivity that is almost independent of the underlying chemistry, in stark contrast to what occurs in crystalline conductors. In this work, we employ an accurately fine-tuned machine learning interatomic potential to construct large-scale molecular dynamics trajectories encompassing hundreds of nanoseconds to obtain statistically converged transport properties. We find that the amorphous state consists of chain fragments of metal-anion tetrahedra of various lenght. By analyzing the residence time of alkali cations migrating around tetrahedrally-coordinated trivalent metal ions, we find that oxygen anions on the metal-anion tetrahedra limit alkali diffusion. By computing the full Einstein expression of the ionic conductivity, we demonstrate that the alkali transference number of these materials is strongly influenced by distinct-particles correlations, while at the same time they are characterized by an alkali Haven ratio close to one, implying that ionic transport is largely dictated by uncorrelated self-diffusion. Finally, by extending this analysis to chemical compositions $AMX_{2.5}\textsf{O}_{0.75}$, spanning different alkaline ($A$ = Li, Na, K), metallic ($M$ = Al, Ga, In), and halogen ($X$ = Cl, Br, I) species, we clarify why the diffusion properties of these materials remain largely insensitive to variations in atomic chemistry.
\end{abstract}
\maketitle
\noindent
\section{introduction}
All solid-state batteries (ASSBs) are a promising next-generation technologies for energy-storage. Their potential advantage over current lithium ion batteries (LIBs) consists in the replacement of the carbonate-based, organic liquid electrolyte in LIBs with an inorganic solid-state electrolyte (SSE) \cite{famprikis_fundamentals_2019, tian_promises_2021}. This replacement has the potential to significantly enhance thermal stability and safety (owing to the non-flammable nature of the SSE), enable higher energy densities (through the possible integration of metallic Li as the anode), and achieve ionic conductivities of up to $\sim$10 mS cm$^{-1}$. Sulfides and oxides have been investigated as SSE, and design rules have been devised to optimize their performances \cite{jun_diffusion_2024,wang_design_2015,jun_lithium_2022,wang_design_2023,wang_design_2023b}. Sulfides can achieve excellent ionic conductivities, up to 30 mS cm$^{-1}$ \cite{kamaya_lithium_2011,seino_sulphide_2014,zhou_new_2019, li_lithium_2023}, but display poor electrochemical stability windows \cite{richards_interface_2016,xiao_understanding_2019}. Oxides improve the electrochemical stability, but at the price of generally inferior conduction properties \cite{zhang_new_2018}. 

In 2018, \citet{asano_solid_2018} suggested halide-based SSE. When synthesized via mechanochemical routes, these materials exhibit ionic conductivities of the order of $\sim$1 mS cm$^{-1}$, while retaining high voltage stability and good deformability \cite{kwak_emerging_2022,liang_metal_2021,yu_design_2023,wang_prospects_2022}. Besides purely halide-based electrolytes, oxyhalides,  mixed-anion conductors that incorporate oxygen within the anion framework, are also being considered \cite{tanaka_new_2023}. Crystalline oxyhalides have been shown to achieve remarkably high ionic conductivities, up to $\sim$10 mS cm$^{-1}$ in lithium-based Nb/Ta oxychlorides, thus rivaling the performances of sulfides \cite{jun_exploring_2025, singh_critical_2024,zhao_anion_2025}. In this context, \citet{dai_inorganic_2023} reported \textit{A}AlCl$_{2.5}$O$_{0.75}$ (\textit{A} = Li$^+$, Na$^+$) oxyhalides, characterized by an amorphous structure. These materials are particularly appealing because they combine: (i) superionic conductivities: $\s_{\textsf{Li}^+}\sim1.5$ mS cm$^{-1}$ and $\s_{\textsf{Na}^+}\sim1.3$ mS cm$^{-1}$; (ii) good chemical stability and mechanical properties, enabling an intimate contact of the electrolyte with high voltage cathodes; (iii) synthesis attained through conventional chemical reactions that bypasses the use of high energy ball-milling, thus paving the way for large scale synthesis. 

Since then, a substantial amount of experimental studies have provided convincing evidence that the ionic conductivity of these materials does not seem to depend significantly on their underlying chemistry \cite{hu_cost-effective_2023,zhang_family_2023,lin_family_2024,li_amorphous_2023,zhang_amorphous_2024,li_cation-anion-engineering_nodate,wu_coordination-disorder_2025,yue_universal_2025}, and is weakly affected by the ionic substitution of both alkali (Li, Na, Ag), metal cation (Al, Zr, Hf, Nb, Ta) and anions (Cl, Br, I) \cite{zhang_universal_2024,wang_oxychloride_nodate}. This behavior is radically different from that of crystalline conductors, where alkaline conductivity has a critical dependence on chemical composition \cite{ong_phase_2012,wang_design_2023b}, especially when Li is substituted by Na or K \cite{kwak_emerging_2022,richards_design_2016}. However, although many studies focused on tuning the oxygen concentration in order to optimize various properties \cite{yang_atomic_2025,kim_oxygen-tuned_2025,hussain_exploring_2024}, the relative independence of $\s$ upon chemistry -- which is highly relevant from both fundamental and applied perspectives -- has not been theoretically addressed yet. Indeed, to the best of our knowledge, computational modeling has so far mainly considered specific chemistries (e.g. Li-Al-Cl-O), without a systematic comparison between different systems. In addition, with the exception of Ref. \cite{kang_non-monotonic_nodate}, computational studies have been performed using \textit{ab initio} molecular dynamics (MD) -- thus using relatively small simulation cells containing $\sim$100 atoms -- which does not allow one to investigate long-range structural features in amorphous materials \cite{gupta_what_2023,grasselli_investigating_2022}, nor to accurately estimate the distinct-particle correlations that contribute to the Green-Kubo ionic conductivity, as this requires extended simulation time scale and sufficient statistical sampling \cite{marcolongo_ionic_2017, fong_ion_2021}. 

In this paper, we present a systematic comparison of amorphous oxyhalides conductors at the $AMX_{2.5}\textsf{O}_{0.75}$ composition. We focus on isovalent chemical substitutions: $A$ as monovalent alkali metal (Li$^{+}$, Na$^{+}$ and K$^{+}$); $M$ as trivalent metal (Al$^{3+}$, Ga$^{3+}$ and In$^{3+}$) and $X$ as a monovalent halide ion (Cl$^-$, Br$^-$ and I$^-$). This systematic chemical modification allows us to unambiguously elucidate the role of chemistry in controlling alkali diffusion.
%, thus avoiding the  comparison of phases with different ionic concentrations; (ii) maintain the same local environment around the trivalent metal cation (tetrahedral atomic coordination within the first shell), thus removing indirect structural effects due to different polyhedral coordination. 
By leveraging accurately fine-tuned CHGNet machine learning interatomic potential (MLIP) \cite{deng_chgnet_2023}, we perform large-scale molecular dynamics (MD) simulations spanning hundreds of nanoseconds in supercells containing thousands of atoms, so that the configurational space of the amorphous phase can be properly sampled. 

We find that the microscopic structure of all investigated materials is characterized by a network of interconnected tetrahedra of trivalent metals coordinated by four halogen/oxygen anions: $[M_m X_n\textsf{O}_\ell]^{3m-n-2\ell}$. These tetrahedra form extended complexes (up to 20 units), and alkali ions diffuse through the interstitial space of this network. By deconvoluting the residence times of alkali cations near metal centers with different chemical coordination, we quantitatively explain why elevated oxygen content reduces alkali diffusion. In addition to the self-particle contributions, we explicitly evaluate the distinct-particle correlations contained within the full Einstein expression of the ionic conductivity, and show that these contributions sizably impact the resulting transference numbers. Finally, the alkali Haven ratio is evaluated to be near unity across all amorphous conductors studied, indicating that -- in contrast to crystalline conductors \cite{marcolongo_ionic_2017} -- alkali transport here is dominated by uncorrelated single-particle diffusion. 

Our results rationalize the microscopic mechanisms that govern the universal diffusion properties across these amorphous conductors, and suggests that this topologically-disordered structural framework can offer concrete alternatives for alkaline non-lithium-ion solid-state batteries.  

\section{Results}
%LiAlCl$_4$ and NaAlCl$_4$ are crystalline, electronically insulating ionic solids, with melting temperatures of 419~K and 430~K, respectively. The addition of oxygen to their molten phase ($A$AlCl$_{4-2x}$O$_x$, $A=$ Li, Na) produces a dramatic increase in viscosity, and at room temperature it boosts the ionic conductivity by three orders of magnitude \cite{dai_inorganic_2023}. In our study, we thus address $AMX_4$ ($A=$ Li, Na, K; $M=$ Al, Ga, In; $X=$ Cl, Br, I) as crystalline precursors. We consider the LiAlCl$_4$ composition and change one element at a time to isolate the effect of single-ion substitution. This target chemical space is realistic because many of these crystals (e.g. LiAlCl$_4$, NaAlCl$_4$, KAlCl$_4$, LiGaCl$_4$, or their combination like LiGaBr$_4$, NaInBr$_4$, KGaI$_4$) are experimentally synthesizable and share similar structural and thermodynamic properties: trivalent metals $M$ that are 4-fold coordinated with halide anions, which are linked through corner- or edge-sharing alkali-polyhedron in monoclinic or orthorhombic space groups \cite{jain_commentary_2013} (see the Experimental section for the definition of all the crystal structures above mentioned and used in this work).

%They also demonstrate similar melting temperatures: 419 K and 430 K for LiAlCl$4$ and NaAlCl$4$, respectively (as determined experimentally \cite{dai_inorganic_2023}), and in the range 450~K -- 550~K (as predicted by machine learning models \cite{hong_melting_2022}). 

The amorphous structural models were obtained using a melt-and-quench approach: The crystalline structures of $AMX_4$ ($A=$ Li, Na, K; $M=$ Al, Ga, In; $X=$ Cl, Br, I -- see Experimental section for the description of all the crystal structures) were melted by heating to 1000~K and equilibrated at the same temperature. Subsequently, oxygen incorporation is implemented through chemical substitution Cl$_4\rightarrow$Cl$_{2.5}$O$_{0.75}$, by placing oxygen atoms on chlorine sites. The anion substitution is not random, but performed by enumerating possible O/Cl orderings and identifying the structure with the lowest Ewald energy. The resulting oxygen-doped structures are then subjected to an additional equilibration at $T=1000$~K; finally, they are cooled and relaxed -- with respect to both atomic positions and cell parameters -- to a local energy minimum. 

MD production runs have been performed using a fine-tuned version of the universal neural-network interatomic potential CHGNet \cite{deng_chgnet_2023}. The training dataset was constructed using the active-learning approach as implemented in the Vienna Ab initio Software Package (VASP) \cite{jinnouchi_--fly_2019,jinnouchi_descriptors_2020,jinnouchi_phase_2019} (more information on the fine-tuning methodology can be found in the Experimental section). MD trajectories were collected from 3–7 ns simulations performed in a cubic-like cell containing $\sim$1300 atoms (see Fig. \ref{fig::amorphous_structure}d).

%MD production runs have been performed using the universal neural-network interatomic potential CHGNet \cite{deng_chgnet_2023}, which was fine-tuned with a combination of oxygen-doped amorphous structures ($AMX_{2.5}$O$_{0.75}$) and crystalline phases (oxides like Li$_2$O and Al$_2$O$_3$, and chlorides LiAlCl$_4$) in order to maximize the structural diversity of the training dataset (more information can be found in SI). The training dataset was constructed using the active-learning approach as implemented in the Vienna Ab initio Software Package (VASP) \cite{jinnouchi_--fly_2019,jinnouchi_descriptors_2020,jinnouchi_phase_2019}, which is based on Bayesian inference to construct on-the-fly a machine learning potential that predicts energies, force and stresses along the MD run. This methodology allowed us to efficiently explore the Born-Oppenheimer potential energy surface while retaining first-principles accuracy: only \textit{ab initio} energies, forces and stresses have been used for fine-tuning the neural network potential CHGNet, which was employed for production runs. MD trajectories have been collected from 3--5 ns-long simulations using a $\sim$1300 atoms cubic-like box.

\subsection{Structural model of the amorphous phase}
\begin{figure*}[t]
\centering
\includegraphics[width=1.0\textwidth]{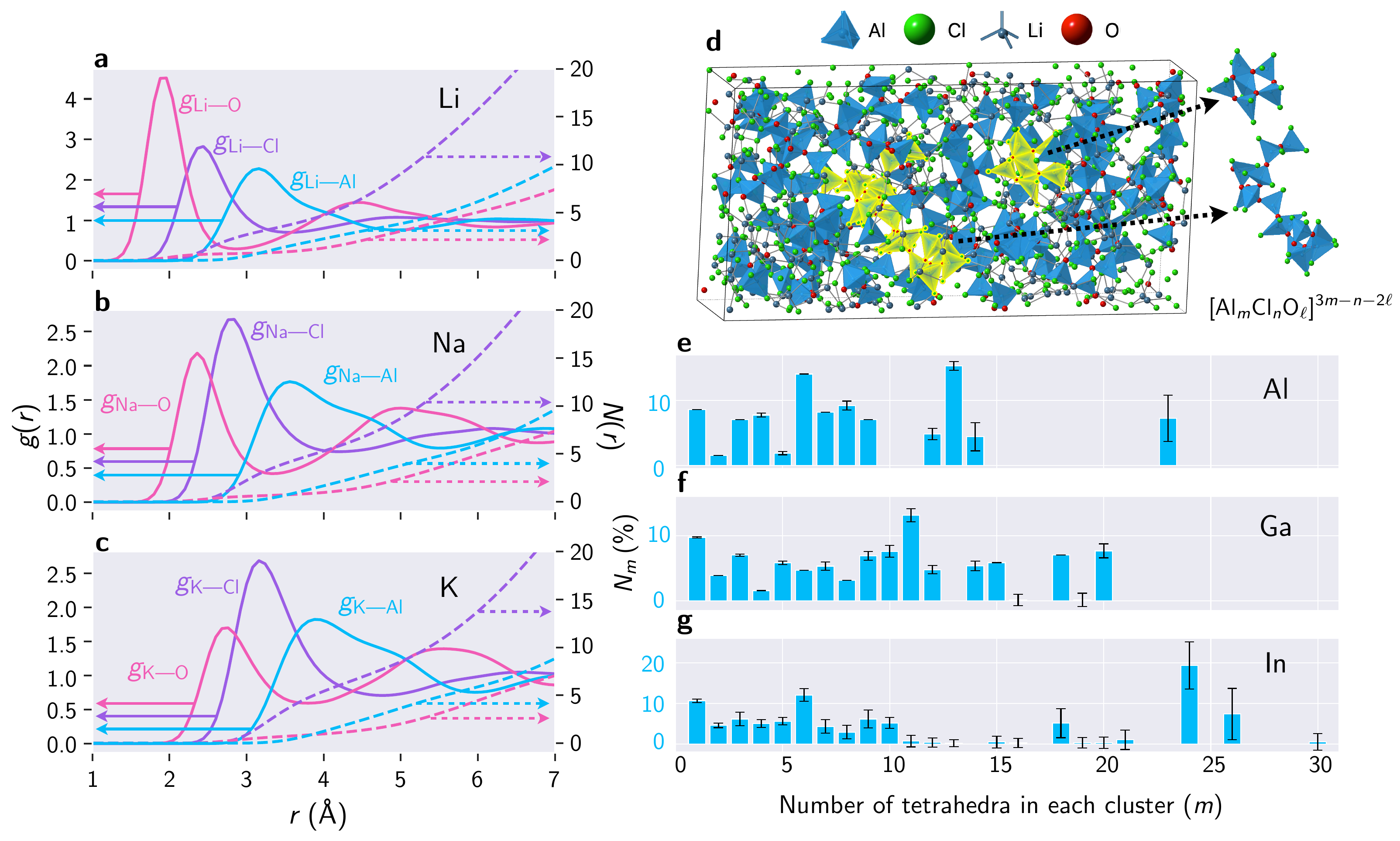}
\caption{\justifying \textsf{\textbf{(a--c)} Radial distribution function (RDF) $g_{A-B}(r)$ (left scale, solid line) and coordination number $N_{A-B}(r)$ (right scale, dashed line) as a function of the distance of the alkali $A=$ Li (a), Na (b), K (c) from the $B$ atom ($B=$ O [pink], Cl [purple], Al [light blue]). \textbf{(d)} Snapshot of the equilibrated MD trajectory of LiAlCl$_{2.5}$O$_{0.75}$; inset shows examples of [Al$_m$Cl$_n$O$_\ell$]$^{3m-n-2\ell}$ complexes. \textbf{(e--g)} Distribution of tetrahedral cluster sizes, where cluster size is defined as the number of tetrahedra contained in each cluster. The y-axis reports the population fraction (in \%). Results are shown for Li$M$Cl$_{2.5}$O$_{0.75}$ ($M=$ Al, Ga, In). Error bars represent the standard deviation $\Delta_m$ (defined in the main text).  }  \label{fig::amorphous_structure}} 
\end{figure*} 
\begin{table}[]
    \centering
\renewcommand{\arraystretch}{1.}
    \setlength\tabcolsep{0.3in}
\caption{\justifying\textsf{Calculated $N_{A-B}(R^*)$ from $g_{A-B}(r)$ displayed in Fig. \ref{fig::amorphous_structure}, where $R^*$ is the minimum point of $g_{A-B}(r)$ after its first maximum. For $B=$ Al, $R^\ast$ is the inflection point of $g_{A-\textsf{Al}}(r)$ after its first peak.}\label{tab::coordination_number}}
\begin{tabular}{cccc}
        \toprule
        Alkali&Cl&O&Al\\
        \midrule
        Li & 3.58 &0.69&  1.41\\
        Na & 5.42 &0.62&  1.77\\
        K  & 7.21 &0.78&  1.92\\
        \bottomrule
    \end{tabular}
\end{table}
In this section, we characterize the distributions of bonds and atomic environments in the amorphous structures obtained from the melt-and-quench simulation. Fig. \ref{fig::amorphous_structure} (panels a--c) shows the radial distribution functions (RDF) $g_{A-B}(r)$ of $A$AlCl$_{2.5}$O$_{0.75}$ between an alkali $A$ ($A=$ Li, Na, K) and the specie $B$ ($B=$ O, Cl, Al). The right axis shows the volume integral of $g_{A-B}(r)$, yielding $N_{A-B}(R)$, which gives the number of $B$ atoms within a distance $R$ of the alkali ion $A$ (see Experimental section for their definitions). In addition to the expected shift to larger distances—reflecting the increase in alkali ionic radius—Fig.~\ref{fig::amorphous_structure}(a–c) shows that the ratio between $\max(g_{A\text{-}\textsf{O}})$ and $\max(g_{A\text{-}\textsf{Cl}})$ decreases as the alkali radius increases. By defining $R^*$ as the first distance at which the minimum of $g_{A-B}(r)$ occurs after its first maximum, $N_{A-B}(R^*)$ quantifies the number of $B$ ions in the first (anions) and second (metal cations) coordination shells of $A$. The calculated values of $N_{A-B}(R^*)$ are reported in Table \ref{tab::coordination_number}. They show that the total coordination number of $A$ with anions increases from Li to K, consistent with Pauling's radius-ratio rule. The total average alkali-anion coordination numbers for Li, Na, and K are 4.27, 6.04, and 7.89, respectively. As shown in Table \ref{tab::coordination_number}, the increase of the coordination number from Li to Na and K is mostly due to an increase of Cl atoms in the local environment of the alkali, as the oxygen coordination remains almost the same.

\begin{figure*}[t] 
\centering
\includegraphics[width=0.80\textwidth]{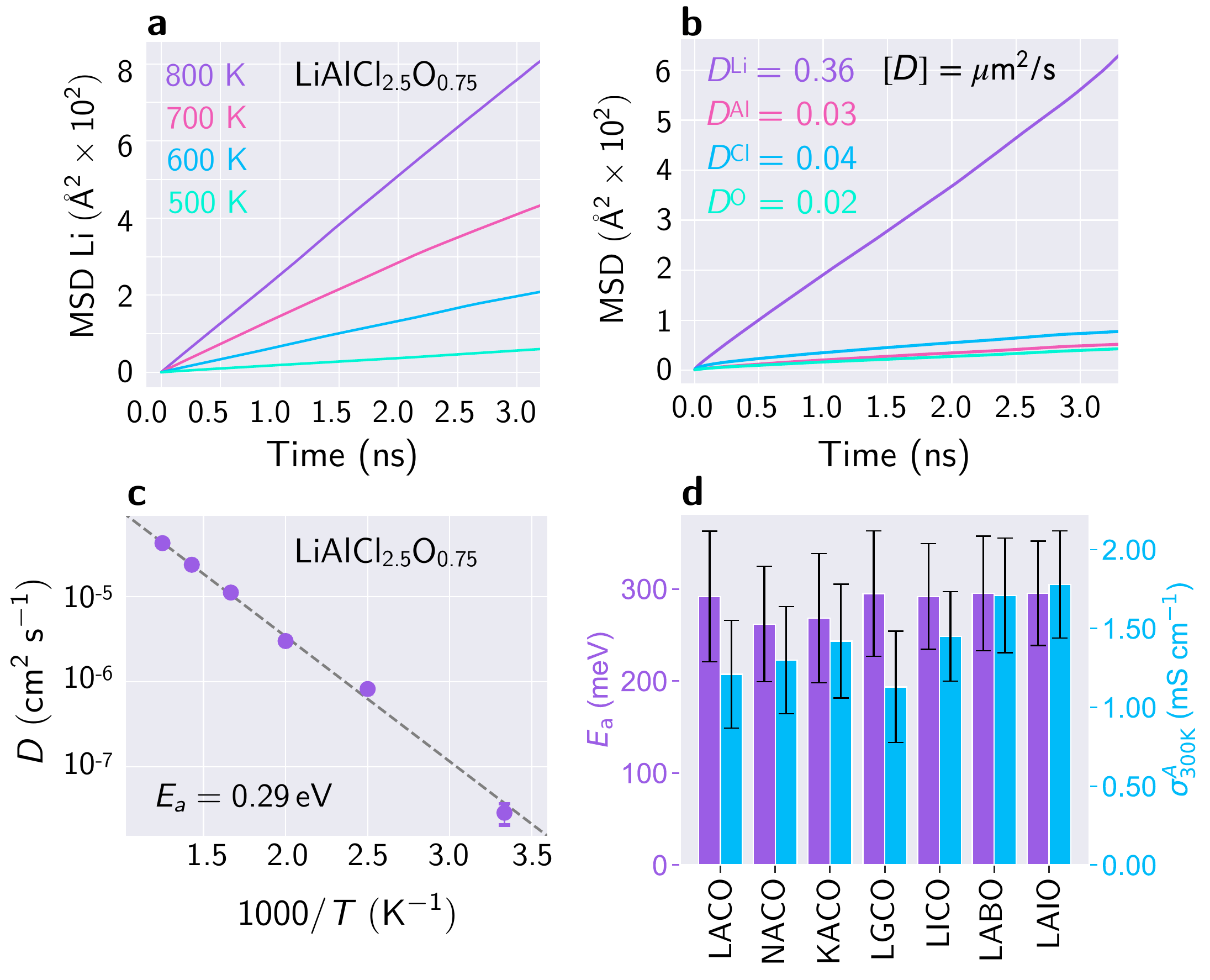}
\caption{\justifying \textsf{\textbf{(a)} Mean square displacement (MSD) of Li atoms in LiAlCl$_{2.5}$O$_{0.75}$ at different temperatures:  500 K (turquoise), 600 K (blue), 700 K (pink) and 800 K (purple). \textbf{(b)} MSD of the different species (Li [purple], Al [pink], Cl [blue], O [turquoise]) of LiAlCl$_{2.5}$O$_{0.75}$; the insets report the associated (tracer) diffusion coefficients $D$ at $T=500$ K in a color-coded manner. \textbf{(c)} Arrhenius plot of $D$ as a function of the inverse temperature for LiAlCl$_{2.5}$O$_{0.75}$; the dashed line is the linear fit used to extract the activation energy. \textbf{(a)} Barplot displaying the extracted activation energies ($E_\textsf{a}$) (left scale, purple) the and calculated conductivities of alkali cations at 300 K using the Nernst-Einstein relation ($\sigma^A_{300\textsf{K}}$) (right scale, blue) of the $AMX_{2.5}$O$_{0.75}$ family of conductors.} \label{fig::msd}}
\end{figure*} 

Fig. \ref{fig::amorphous_structure}d displays a snapshot of the amorphous phase of LiAlCl$_{2.5}$O$_{0.75}$, which exhibits extended complexes of metal-anion (\textit{M}$-$\textit{X}) tetrahedra. We define a \emph{complex} as a connected network of \textit{M}$-$\textit{X} polyhedra, which can extend in the space both linearly (i.e. chain-like), or with a planar geometry (see inset of Fig. \ref{fig::amorphous_structure}d). The cutoff distance used to determine whether two metal cations are considered part of a complex corresponds to the first minimum of $g_{M-M}(r)$ (reported in Fig. \ref{fig::g_r_metal-metal}). In these complexes, oxygen atoms act as bridges between different tetrahedra (100\% of O anions are always found to be bonded to metal cations, as shown in Fig. \ref{fig::bound_anions}.), in agreement with previous experimental and theoretical findings \cite{dai_inorganic_2023,yang_atomic_2025}.  Fig.~\ref{fig::amorphous_structure}(e--g) shows the percentage of tetrahedral units ($N_m$) that are part of complexes containing $m$ metal cations. Data was collected from snapshots of MD simulations of $\sim$1 ns at $T=400$ K, with a sampling interval of 20 ps. Results are shown for Li$M$Cl$_{2.5}$O$_{0.75}$, with $M =$ Al, Ga, In (a similar plot for different halogens is shown in Fig. \ref{fig::cluster_halogens}). The associated standard deviations, $\Delta_m$, are indicated by error bars. These quantities are computed as  
${N}_m = \frac{m}{N_\textsf{snapshot}} \sum_{j=1}^{N_\textsf{snapshot}}N_m(j)$  and $\Delta_m = \frac{m}{N_\textsf{snapshot}-1}\sqrt{\sum_{j=1}^{N_\textsf{snapshot}} \big(N_m(j)-{N}_m/m\big)^2}$
where $N_m(j)$ is the number of complexes containing $m$ trivalent metals observed in the $j$-th MD snapshot (by construction, $\sum_m {N}_m
=N_{\textsf{metal}}$ -- which was numerically verified). Fig. \ref{fig::amorphous_structure} (e--g) shows that only $\sim$10\% of $M$ cations form isolated polyhedra, while the rest form extended tetrahedral complexes, comprising tens of units. 

The Al-Cl-O geometries found from our simulations are consistent with Zhang et al. \cite{zhang_universal_2024}, who observed a coexistence of AlCl$_4^-$ species and $[\textsf{Al}_m\textsf{Cl}_n\textsf{O}_\ell]^{3m-n-2\ell}$ clusters from both nuclear magnetic resonance (NMR) measurements and pair distribution function (PDF) analysis. In addition, the measured bond lengths of Al--Cl ($\sim$1.75 \AA) and Al--O ($\sim$2.15 \AA) from X-ray and neutron-based PDF of Wang and coworkers \cite{wang_oxychloride_nodate} closely match our calculated Al--Cl and Al--O peaks in the corresponding RDF (see Fig. \ref{fig::residence_time_analysis}d).  
%The complexes' size fluctuations, represented by their error bars, increase from Al$\rightarrow$Ga$\rightarrow$In. With equal cation valence charge, this is likely due to the lower polarizing power of metals with larger size  

%, metals with larger size (In and Ga) have lower polarizing power, leading to larger complexes-lengths' fluctuations across the MD simulation.   

\begin{figure}[t]
\centering
\includegraphics[width=0.48\textwidth]{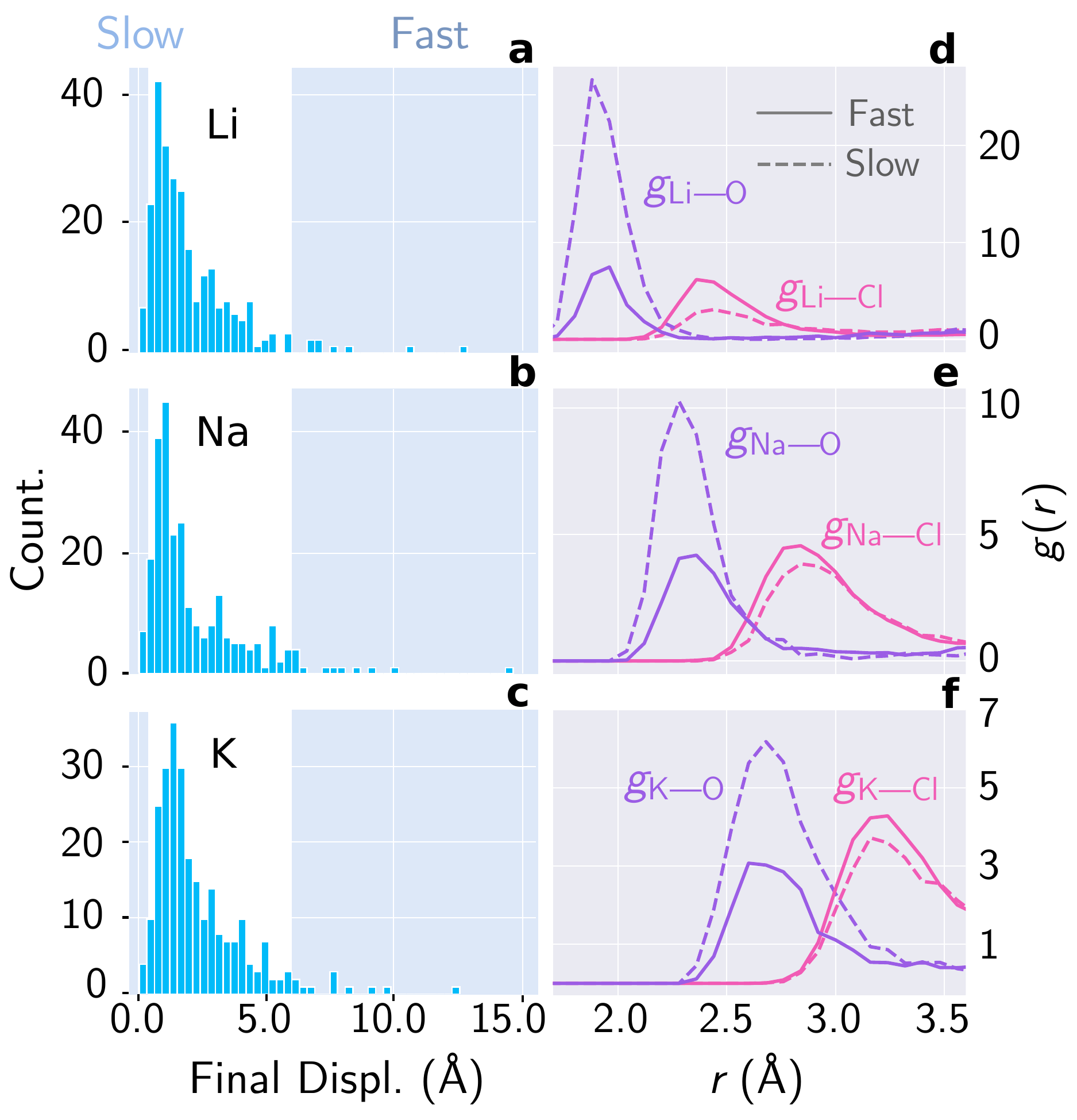}
\caption{\justifying \textsf{\textbf{(a--c)}: Histograms of the alkali ion displacements after 4 ns simulation time for Li (a), Na (b), and K (c). Highlighted areas define the 5\% fastest- and 5\% slowest-moving ions, respectively. \textbf{(d--f)}: radial distribution functions between alkali (Li [d], Na [e], K [f]) and Cl (pink) or O (purple), for the fast- (continuous lines) and the slow-moving (dashed lines) ions.} \label{fig::slow_fast}}
\end{figure} 
\subsection{Diffusion properties and residence time analysis}
\begin{figure*}[t]
\centering
\includegraphics[width=0.99\textwidth]{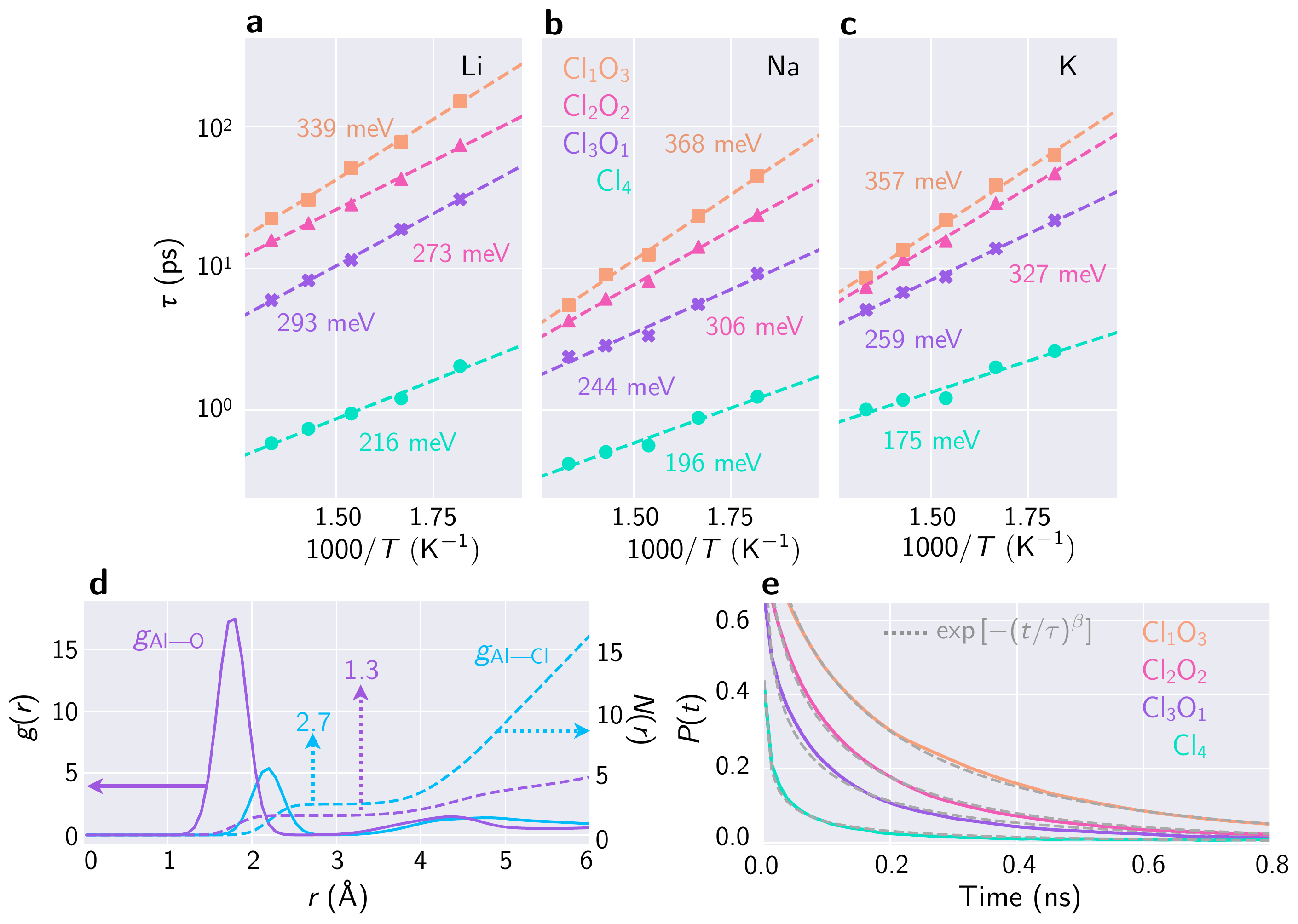}
\caption{\justifying \textbf{(a--c)} Residence times $\tau$ extracted from the exponential fit of $P(t)$ as a function of the inverse temperature for Li \textbf{(a)}, Na \textbf{(b)} and K \textbf{(c)}; dashed lines are linear fits, and the corresponding slope is proportional to the corresponding activation energies $\varepsilon_\textsf{a}$, which are reported in a color-coded manner. \textsf{\textbf{(d)} Radial distribution function (continuous lines, left axis) and coordination number (dashed lines, right axis) as a function of distance for Al-Cl pair (light blue) and Al-O pair (purple) in LiAlCl$_{2.5}$O$_{0.75}$. \textbf{(e)} calculated $P(t)$ as a function of time for an MD simulation at $T=500$ K of LiAlCl$_{2.5}$O$_{0.75}$ for Li near different Al-anion tetrahedra; dashed grey lines are the corresponding exponential fits. } \label{fig::residence_time_analysis}}
\end{figure*} 
Fig. \ref{fig::msd} displays the calculated diffusion coefficients of the alkali ions in  $AMX_{2.5}$O$_{0.75}$ amorphous oxyhalides. The mean square displacements (MSD) in LiAlCl$_{2.5}$O$_{0.75}$ (LACO) are shown in Fig. \ref{fig::msd}a for Li at various $T$, and in Fig. \ref{fig::msd}b for all species of LACO at $T=500$ K. Their linear increase with time indicates that a diffusive regime has been reached in the simulations. Fig. \ref{fig::msd}c shows the Arrhenius plot LACO (and in Fig. \ref{fig::all_arrhenius} for all the other materials). The linearity of this Arrhenius plot confirms that no phase transition or change in mechanism occurs within the analyzed temperature range. Fig. \ref{fig::msd}d shows the calculated room temperature conductivities ($\sigma_{300\textsf{K}}$) and the activation energies extracted from the Arrhenius plot ($E_{\textsf{a}}$). All studied oxyhalides exhibit $\sigma_{300\textsf{K}}$ of the order of $\sim1$ mS cm$^{-1}$ and activation energies of the order of 0.3 eV, which agree well with the measurements of Dai et. al \cite{dai_inorganic_2023} ($\sigma^{\textsf{LACO}}_{300\,\textsf{K}}=1.5$ mS cm$^{-1}$ and $E_{\textsf{a}}^{\textsf{LACO}}=0.33$ eV) at the same anion composition. The excellent agreement with experimental results lends confidence to both the reliability of the model amorphous structures and their microscopic mechanisms governing the calculated diffusion properties. 

To understand the nature of the alkali diffusion, we performed an analysis similar to \citet{yang_fast_nodate}, where we examine the local environment of Li, Na, and K cations after we separate the alkali cations into fast- and slow-moving atoms within a low-temperature simulation ($T=300$ K). Fast- and slow-diffusing alkali refer to the respective atoms with total displacement $|\Delta\boldsymbol{r}|>0.95|\Delta\boldsymbol{r}_{\textsf{max}}|$ and $|\Delta\boldsymbol{r}|<0.05|\Delta\boldsymbol{r}_{\textsf{max}}|$, respectively, where $|\Delta\boldsymbol{r}_{\textsf{max}}|$ is the maximum alkali displacement observed at the end of a 4 ns long MD simulation. In Fig. \ref{fig::slow_fast}(a--c), histograms of the displacements are reported, where the separations between fast- and slow-moving alkali are highlighted, and Fig. \ref{fig::slow_fast}(d--f) shows the RDF between the fast- and slow-moving alkali and the anions. The results of Fig. \ref{fig::slow_fast} indicate that slow-diffusing ions tend to have a more pronounced amount of oxygen in the first coordination shell. It should be noted that in the limit of infinite simulation time, the amount of oxygen concentration around every alkali should be the same, and a separation between fast- and slow-moving ions is ill-defined because of the underlying assumption of ergodicity. However, this analysis still suggests that, on relatively short time scales (nanoseconds), alkali ions trapped in local oxygen-rich environments diffuse more sluggishly.

To make this characterization more rigorous, we performed a residence time analysis that allows us to unambiguously extract the residence time of an alkali cation when it migrates away from a predefined tetrahedral environment. To this aim, and following established approaches  for modeling liquid electrolytes \cite{borodin_litfsi_2006,borodin_mechanism_2006,self_transport_2019}, we introduce the quantity:
    \begin{gather}
    P(t) =\frac{1}{N_{t_i}}\sum_{i=1}^{N_{t_i}}\frac{\sum_{nm}[Q(t_i+t)\circ Q(t_i)]_{nm}}{\sum_{nm}Q(t_i)_{nm}};\\
    Q_{nm}=\begin{cases}1\quad\textsf{distance}(A_n,M_m)<\bar{d}\\ 0 \qquad \textsf{otherwise}\end{cases}
\end{gather}
where $\circ$ denotes the Hadamard product (element-wise product of the $Q$ matrices), and $\bar{d}$ is a cutoff distance corresponding to the inflection point of $g_{A-\textsf{Al}}(r)$ after its first maximum (Fig. \ref{fig::amorphous_structure}(a--c)). 
%$P(t)$ is an autocorrelation function that rapidly decays to zero. 
\(P(t)\) can be interpreted as the survival probability of
alkali--polyanion coordination pairs, i.e., the probability that an alkali ion and a metal tetrahedron initially
coordinated at a reference time \(t_i\) remains within the first coordination
shell after a time lag \(t\).
As such, \(P(t)\) directly quantifies the temporal stability and characteristic
lifetime of local solvation or coordination environments between alkali and polyanions within the eletrolyte.
A slow decay of \(P(t)\) indicates long-lived, persistent coordination motifs,
whereas a rapid decay reflects frequent exchange processes and highly dynamic
local structures.
The residence time $\tau$ is extracted by fitting the $P(t)$ curves with a stretched exponential function: $P(t)\sim\exp[-(t/\tau)^\beta]$. 

Focusing on $A\textsf{AlCl}_{2.5}\textsf{O}_{0.75}$ ($A=$ Li, Na, K), we track the residence time of an alkali located in proximity of an Al tetrahedron, differentiating whether the latter is
coordinated with: Cl$_4$,  Cl$_3$O$_1$, Cl$_2$O$_2$ and Cl$_1$O$_3$. To verify 4-fold Al coordination, Fig. 4d shows the RDF $g_{\textsf{Al}-\textsf{anions}}$ and the integrated coordination number $N_{\textsf{Al}-\textsf{anions}}$. The isolated oxygen and chlorine peaks, together with their integral, confirm that Al is always 4-fold coordinated with the anions ($N_{\textsf{Al}-\textsf{O}}(r=2.6 \,\textsf{\AA})+N_{\textsf{Al}-\textsf{Cl}}(r=2.9 \,\textsf{\AA} )=4.07$). Fig. \ref{fig::residence_time_analysis}e shows an example of the calculated $P(t)$ at $T=550$ K for LiAlCl$_{2.5}$O$_{0.75}$, and the corresponding exponential fits, from which we extracted the parameters $\beta$ and $\tau$.  The exponents $\beta$ are found to range between 0.25 and 0.65 (see Fig. \ref{fig::beta}). Values of $\beta$ in the range $0<\beta<1$ are widespread in the context of electronic and molecular systems, from glass relaxation processes \cite{welch_dynamics_2013} to charge diffusion in amorphous semiconductors \cite{murayamaand_monte_1992}, and can be rationalised in terms of microscopic theoretical models \cite{phillips_stretched_1996}. Figs. \ref{fig::residence_time_analysis}(a--c) shows the temperature-dependent $\tau=\tau(T)$. By assuming an exponential Arrhenius-like $T$-dependence of the form $\tau(T)\propto\exp{\big[\varepsilon_\textsf{a}/(k_\textsf{B}T)\big]}$, we can extract the activation energies $\varepsilon_\textsf{a}$. The resulting $\varepsilon_\textsf{a}$ are reported color-coded in the same plot and are found to be of the same order of magnitude as the activation energy $E_\textsf{a}$ obtained from fitting the diffusion coefficients in Fig. \ref{fig::msd}d (i.e. between 0.2 and 0.4 eV). The activation energies are systematically the smallest for 4-fold Cl coordination, when no oxygen is present in the tetrahedral environment. This is consistent with the previously established understanding that larger anions with more diffuse and lower charge bind the alkali cation less \cite{jun_diffusion_2024, Canepa2017}

\subsection{Atomistic correlations underlying ionic conductivity}
\begin{figure*}[t]
\centering
\includegraphics[width=1.0\textwidth]{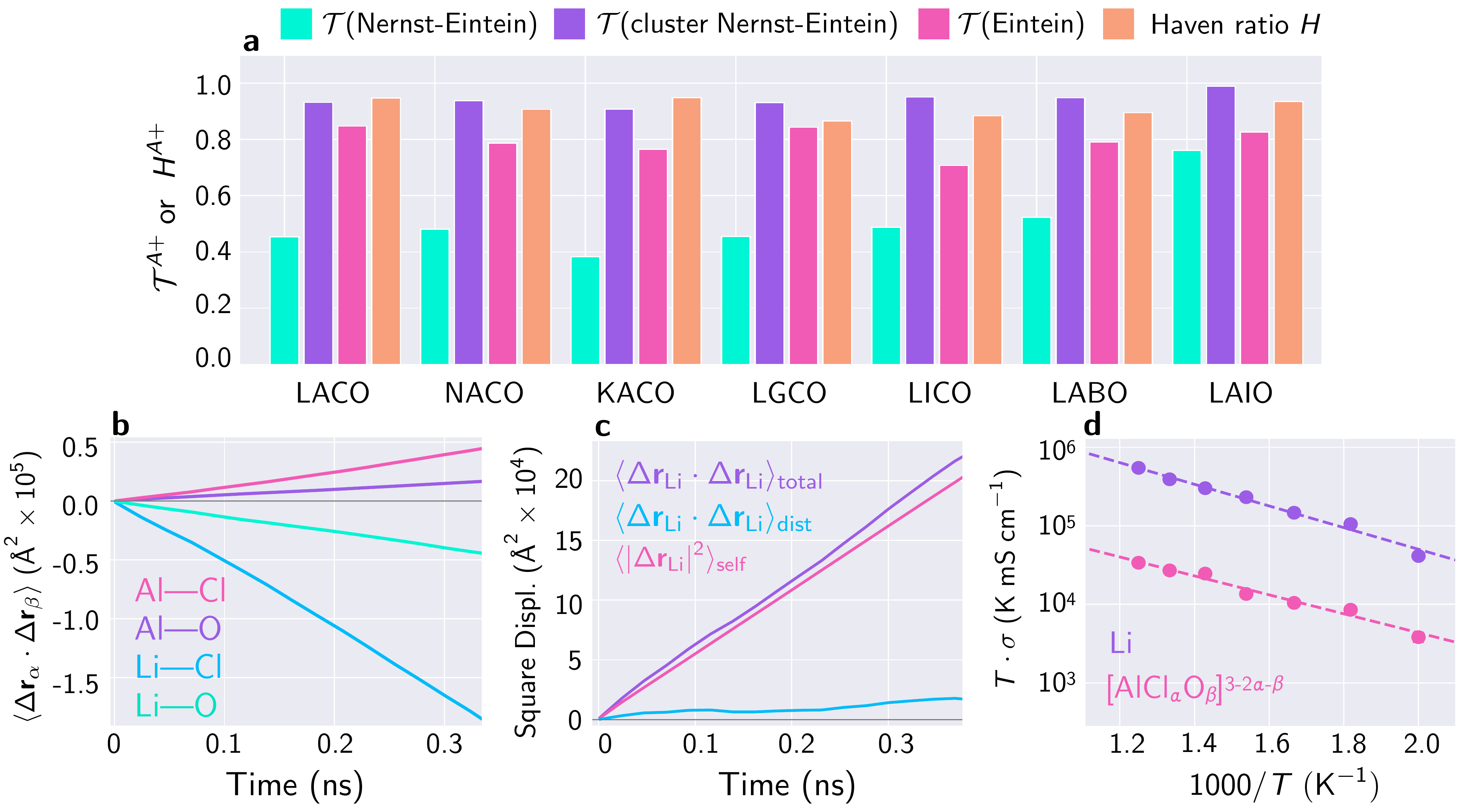}
\caption{\justifying \textbf{(a)} Alkali transference numbers $\mathcal{T}^{A+}$ calculated with Nernst-Einstein (NE) approximation (light blue), cluster NE (purple) and from the Einstein expression (pink) for the $AMX_{2.5}\textsf{O}_{0.75}$ family of compounds, together with alkali the Haven ratio $H^{A+}$ (orange) extracted from MD simulation at $T=750$ K. \textbf{(b)} Displacement-displacement correlation function (DDCF) of LiAlCl$_{2.5}$O$_{0.75}$ between Li-anions and Al-anions. \textbf{(c)} Li-Li DDCF (purple line), mean square displacement (MSD) multiplied by $N^\textsf{Li}$ (pink line), and distinct part of Li-Li DDCF (light blue line) in LiAlCl$_{2.5}$O$_{0.75}$; the distinct part represents the off-diagonal elements of the Li-Li DDCF. \textbf{(d)} Conductivities (multiplied by their corresponding temperatures) as a function of the inverse temperature of Li ions, calculated with the NE approximation, and of [Al$_m$Cl$_n$O$_\ell$]$^{3m-n-2\ell}$ clusters, evaluated within the cNE approximation. \label{fig::onsager}}
\end{figure*} 
The general and exact form of the ionic conductivity is given by the time integral of the current-current autocorrelation function (Green-Kubo expression) \cite{kubo_statistical-mechanical_1957,grasselli_topological_2019,french_dynamical_2011}. By standard manipulation, it can be expressed in an Einstein form \cite{fong_ion_2021,grasselli_invariance_2021}: 
\begin{align}
    \!\!\!\sigma_\textsf{E}&=\frac{e^2}{6Vk_\textsf{B}T}\lim_{t\rightarrow\infty}\frac{d}{dt}\sum_{\alpha,\beta}^N \sum_{i,j}^{N_\alpha,N_\beta}\!\! Z_\alpha Z_\beta\big\langle\Delta\textbf{r}_i^\alpha(t)\cdot\Delta\textbf{r}_j^\beta(t)\big\rangle\label{eq::Einstein}\\
    &\equiv\frac{e^2}{6Vk_\textsf{B}T}\sum_{\alpha,\beta}^N L_{\alpha\beta}\label{eq::L_terms}
\end{align}
where $\Delta \boldsymbol{r}(t)\equiv\boldsymbol{r}(t)-\boldsymbol{r}(0)$, $\alpha$ indicates the atomic specie of the ion indexed by $i$ and $Z_\alpha$ is its valence charge. This expression of $\sigma_{\textsf{E}}$ contains different terms: those with $\alpha\neq\beta$ are the \emph{inter-species} components, and those with $\alpha=\beta$, that are the \emph{intra-species} parts. Within the intra-species components, we can further identify the "traced", or \emph{self} terms (i.e. when $i=j$ in Eq. (\ref{eq::Einstein})) and the \emph{distinct} contributions ($i\neq j$). We will make use of this terminology hereafter. Eq. (\ref{eq::Einstein}) is generally difficult to calculate with sufficient statistical reliability, as the non-diagonal terms in the summation ($\alpha\neq\beta$ and/or $i\neq j$) converge poorly over conventionally accessible time scales in simulations. For this reason, it is common practice to retain only the diagonal terms (i.e. self self-correlations) in the sum ($\alpha=\beta$ and $i=j$), which exhibit a time convergence significantly faster than the off-diagonal ones. Such an approximation is called the Nernst-Einstein (NE) approximation: $\sigma_\textsf{NE}^\alpha=\frac{(Z_\alpha e)^2}{6Vk_\textsf{B}T}\lim_{t\rightarrow\infty}\frac{d}{dt}\sum_{i=1}^{N_\alpha}  \big\langle\big|\textbf{r}^{\alpha}_i(t)-\textbf{r}^\alpha_i(0)\big|^2\big\rangle\equiv\frac{(Z_\alpha e)^2}{6Vk_\textsf{B}T}\,N^\alpha D_{\textsf{tr}}^\alpha$ and is the standard used in \emph{ab initio} MD studies. $D_\textsf{tr}^\alpha$ is the tracer diffusion coefficient of the atomic specie $\alpha$, and is obtained by fitting the linear region of the mean square displacement (MSD) as a function of time. 

To incorporate the effects of ionic correlation, not captured in the NE approximation but generally present in Eq. (\ref{eq::Einstein}), France-Lanord and Grossmann introduced the cluster Nernst-Einstein (cNE) approximation \cite{france-lanord_correlations_2019}, which accounts for anion-cation correlations in the particular case when anions and cations form well-defined clusters. This approximation is valid when metals form tetrahedral clusters that remain intact across the MD simulation. Fig. \ref{fig::onsager}b shows the displacement-displacement correlation function (DDCF) between cations and anions, which is $\sum_{i,j}\langle\textbf{r}_i^{\textsf{cation}}(t)\cdot\textbf{r}_j^{\textsf{anion}}(t)\rangle$ as defined in Eq. (\ref{eq::L_terms}). The strictly positive slope of DDCF between metals $M$ and anions shows that $M$ tend to drift with their corresponding anion tetrahedral cages. Conversely, the DDCF between alkali $A$ and anions is negative (Fig. \ref{fig::onsager}b), meaning that $A$ diffuses in an anticorrelated manner with respect to the tetrahedral clusters, and -- most importantly -- not coordinated with fixed surrounding anions during their diffusion. 

We adopted a variant of the cNE approximation, which we describe in the following: for each cluster, we: (i) determine the total number of metal-anion tetrahedra it is composed of ($N_M$); (ii) calculate the corresponding oxidation number of this cluster: $\Omega=3N_M-2N_\textsf{O}-N_X$ ($M=$ metal, $X=$ halide); (iii) assign to each of the $N_M$ metal-anion tetrahedra, an oxidation state equal to $\tilde{Z}=\Omega/N_M$, (iv) the geometrical center of each tetrahedra is identified with the position $\textbf{R}$ of the metal cation. The metal–anion contribution to the total ionic conductivity within the cNE framework is therefore given by:
\begin{equation}
    \sigma_{\textsf{cNE}}^{M}=\frac{e^2}{6Vk_{\textsf{B}}T}\lim_{t\rightarrow\infty}\frac{d}{dt}\sum_{i=1}^{N_M}\tilde{Z}_i^2\,\big\langle\big|  \textbf{R}_i(t)-\textbf{R}_i(0)\big|^2\big\rangle
\end{equation}
To account for possible fluctuations in the cluster size (as observed for $M = $ In, Fig. \ref{fig::amorphous_structure} (g)), we determined effective oxidation numbers $\tilde{Z}$ by averaging them over 100 equally-spaced snapshots collected throughout the entire MD trajectory. With these definitions, we can evaluate the transference number $\mathcal{T}$ of the diffusing alkali cations $A$ as: $\mathcal{T}_{\textsf{cNE}}^{A}=\sigma_\textsf{NE}^A/(\sigma_\textsf{NE}^A+\sigma_\textsf{cNE}^M)$. Such expression of $\mathcal{T}_{\textsf{cNE}}^A$ assumes that the alkali diffusion can be adequately described by the tracer diffusion coefficient $D_\textsf{tr}^A$, thereby neglecting alkali-alkali, distinct, intra-particles correlations. This assumption will be justified at the end of this section.

An equivalent definition of $\mathcal{T}^A$ within the NE approximation reads: 
\begin{equation}
    \mathcal{T}^A_\textsf{NE}=\frac{\big(Z^A\big)^2 N^A D^A_\textsf{tr}}{\sum_{\alpha=A,M,X,\textsf{O}}\big(Z^\alpha\big)^2 N^\alpha D^\alpha_\textsf{tr}}.
\end{equation} 
Finally, using the exact expression of Eq. (\ref{eq::Einstein}),
the transference number in the Einstein form is given by \cite{fong_ion_2021}: 
\begin{equation}
    \mathcal{T}^A_\textsf{E}=\frac{L_{A-A}+2L_{A-M}+2L_{A-X}+2L_{\textsf{O}-A}}{\sum_{\alpha} \sum_\beta L_{\alpha\beta}}\label{eq::Einstein_tA}
\end{equation}
Fig. \ref{fig::onsager}a shows the calculated $\mathcal{T}^A$ within the three levels of approximations; each of them are the resulting average of five different temperatures: 600, 650, 700, 750 and 800 K. These results show that the NE significantly underestimates the transference numbers, predicting them to be $\sim$0.5. This result stems from the fact that, although the alkali mobility is substantially larger than that of the halides ions (as shown in Fig. \ref{fig::msd}b), the larger number of $X$ (2.5 times larger than $A$) results in a contribution $N^XD^X_\textsf{tr}$ of the same magnitude as $N^AD^A_\textsf{tr}$. The only exception is LiAlI$_{2.5}$O$_{0.75}$ (LAIO), having $\mathcal{T}_{\textsf{NE}}=0.75$, which originates from a much slower framework mobility as compared to the other materials, likely due to a larger inertia caused by heavier iodine anions. For example, at $T=500$~K, $\sigma_{\textsf{NE}}^{\textsf{non-Li}}(\textsf{LAIO})=34$ mS cm$^{-1}$ and $\sigma_{\textsf{NE}}^{\textsf{non-Li}}(\textsf{LACO})=117$ mS cm$^{-1}$.
%, and the Arrhenius plots of LAIO within the cNE (displayed in Fig. \ref{fig::cNE_series}f), shows that $\sigma_{\textsf{cNE}}^{\textsf{non-Li}}(\textsf{LAIO})/\sigma_{\textsf{NE}}^{\textsf{Li}}(\textsf{LAIO})\approx 10^{-2}$, while for the other materials $\sigma_{\textsf{cNE}}^{\textsf{non-Li}}/\sigma_{\textsf{NE}}^{\textsf{Li}}\approx 10^{-1}$ 
The cNE tends to improve the agreement with the exact (Einstein) results, although it overestimates the transference numbers. The origin of these deviations is evident from Fig. \ref{fig::onsager}d and Fig. \ref{fig::cNE_series}, showing the Arrhenius plots of the cNE conductivity for LACO and for the other materials, respectively. The conductivity associated to the metal-anion tetrahedral complex is one order of magnitude smaller than the alkali ions, thus resulting in $\mathcal{T}_\textsf{cNE}^A$ values close to 1. The calculated values of $\mathcal{T}_{\textsf{E}}$ from the full Einstein expression (Fig. \ref{fig::onsager}a, pink bars) are intermediate between $\mathcal{T}_\textsf{NE}$ and $\mathcal{T}_\textsf{cNE}$ and agree well with the experimental estimates from Ref. \cite{you_facile_2025} (0.98) and Ref. \cite{dai_inorganic_2023} (0.69 for LACO and 0.91 for NACO). This arises from a cancellation of inter-particle contributions (which are tabulated in Table \ref{table::einstein_coeff}), resulting in an overall conductivity that is to the greatest extent governed by alkali diffusion.

It is worth noticing that all the investigated materials, regardless of the level of approximation, exhibit similar transference numbers. This constitutes additional evidence that the microscopic transport mechanisms of these amorphous oxyhalides is universal and encompasses all the investigated compounds.

Finally, we address the role of the alkali-alkali correlations. These are encoded in the off-diagonal (intra-particles, distinct) elements of the DDCF. A quantification of such correlation for the alkali specie $A$ (= Li, Na, K) is captured by the Haven ratio \cite{murch_haven_1982}, which is given by $H^A=N^AD^A_\textsf{tr}/D^A_\sigma$, where $D_\sigma^A$ is \emph{defined} by the relation $\sigma^A_\textsf{E}=\frac{(Z^A e)^2}{6Vk_\textsf{B}T}\,D^A_{\sigma}$. Deviations from unity of $H^A$ imply that distinct, intra-particle correlations between the alkali cations influence the macroscopic conductivity of the electrolyte. Fig. \ref{fig::onsager}c shows the DDCF for LiAlCl$_{2.5}$O$_{0.75}$, and its linearity is confirmed by the same log-log scale plot in Fig. \ref{fig::log-log}  (equivalent plots for the whole series of materials are reported in Fig. \ref{fig::for_haven}). Remarkably, we find that the DDCF of alkali cations is largely determined by self-correlations, while the distinct terms negligibly contribute to $\sigma_\textsf{E}^{A-A}$. This trend is observed for the whole $AMX_{2.5}$O$_{0.75}$ family of conductors, as all of the calculated alkali Haven ratio are found to be close to unity.
%\item We now analyze the microscopic conduction mechanisms of the $AMX_{2.5}\textsf{O}_{0.75}$ family of compounds.
%5\item The ionic conductivity of an electrolyte is an example of the Onsager reciprocal relations, and represents one of the Onsager transport coefficients. 

\section{Discussion}
In this work, we report a systematic computational study of the $A$-$M$-$X$-O family of amorphous oxyhalide superionic conductors at $AMX_{2.5}\textsf{O}_{0.75}$ composition.  %Amorphous structural models of $AMX_{2.5}\textsf{O}_{0.75}$ have been constructed starting from $AMX_4$ crystalline structures, and incorporating oxygen with a melt-and-quench approach. 
%All the MD calculations for production runs have been performed using machine learning interatomic neural-network potentials (CHGNet), which was fine-tuned using an on-the-fly active learning methodology that allowed us an highly accurate and efficient exploration of the phase-space of amorphous structures. Large cells containing thousands of atoms have been employed and ns-long MD simulation have been performed, thanks to which the diffusional properties of the investigated materials have been correctly reproduced. 
We find that the amorphous structures consist  of extended clusters formed by interconnected metal-anion tetrahedra, linked through corner- or edge-sharing oxygen atoms. The alkali cations are found to coordinate mostly with chlorine atoms;  cations with increasing ionic radius (Li$\rightarrow$Na$\rightarrow$K) exhibit higher coordination numbers accomplished by increasing the Cl$^-$ in the first anion shell, while the amount of oxygen remains relatively constant (one O atom in the first coordination shell). The extended clusters of metal-anion tetrahedra are observed in all compositions. 
%Cations with larger radii exhibit wider clusters' size fluctuations, which is likely due to their lower polarizing power due to their larger spatial extension at same valence charge. A similar effect appears also for systems different halide radii (as shown in SI). 

A residence time analysis has shown that alkali diffusing in proximity of aluminum coordinated with chlorine only (Cl$_4$) exhibit residence times that are systematically one order of magnitude smaller than those obtained around oxygen-rich environments (Cl$_{4-n}$O$_n$). The activation barriers $\varepsilon_\textsf{a}$ extracted from the temperature dependence of the residence times exhibit values that 
fall within the same range as the activation energies determined from diffusion coefficients 
%consistent with the activation energies determined from the Arrhenius fit of the diffusion coefficients 
($E_{\textsf{a}}^{D_{\textsf{tr}}}$); this observation tells us that alkali diffusion and their escape times from metal-anion tetrahedra share the same reaction rates, and therefore we can use $\varepsilon_{\textsf{a}}$ as a more chemically-informed proxy to identify the energy barriers associated to different escaping environments. 
%Notably, $\varepsilon_\textsf{a}$ corresponding to Al coordinated with at least one O [$\varepsilon_\textsf{a}$(Cl$_{4rates-n}$O$_n$)]  $E_{\textsf{a}}^{D_{\textsf{tr}}}$, and have systematically larger values than $\varepsilon_\textsf{a}$(Cl$_4$). 
To this aim, we notice that (i) Cl$_{3}$O$_1$ tetrahedra are the most common coordination environment for $M$ (as confirmed from $N^{\textsf{1st\,shell}}_{\textsf{Al}-\textsf{O}}=1.3$, in Fig. \ref{fig::residence_time_analysis}d), and (ii) the activation energies for escaping the Cl$_3$O$_1$ tetrahedra mostly match the ones obtained from the diffusion coefficients: $E_{\textsf{a}}^{D_{\textsf{tr}}}/\varepsilon_{\textsf{a}}^{\textsf{Cl}_{3}\textsf{O}_1}$ are 292/293, 262/244, 268/259 meV for Li, Na and K, respectively. Cl$_4$ tetrehedra have a lower escape time; however, they are unlikely to be percolating and hence do not transfer their low activation energy for alkali hopping to the macroscopic diffusion. This realization clarifies that oxygen substitution for Cl$^-$—despite being necessary for stabilization of the amorphous phase \cite{dai_inorganic_2023}—represents the principal bottleneck to faster ion diffusion. In order to increase alklai conductivity, it may be necessary to limit oxygen content or to implement other amorphization strategies such as the use of divalent anions with lower electronegativity such as sulfur \cite{Ogbolu2025}, selenium or polyanion units \cite{Tan2025}.

%, and it also suggests that the ionic conductivity of these materials could be enhanced by addressing the issue of the divalent anion substitution for Cl$^-$, which appears to be needed for the thermodynamical stabilization of the amorphous phase \cite{dai_inorganic_2023}, but also constitute the most serious bottleneck for even faster ion diffusion. For example, implementing divalent anions with lower electronegativity such as sulfur (for which promising attempts have been made in \cite{dai_inorganic_2023}), selenium or polyanion units \cite{Tan2025} could be possible strategies.

Our data enables a more complete and general picture of the structure and the alkali migration in these materials to emerge. The displacement-displacement correlation function (Fig. \ref{fig::onsager}a) showed that aluminum atoms tend to drift coordinated in the corresponding anion tetrahedral cages. Their diffusion is, however, one order of magnitude slower than that of alkali atoms, which themselves exhibit anti-correlation with respect to such polyanions. 
%The analysis of distinct particles correlations among alkali cations reveals an Haven ratio close to one for all the investigated oxyhalides. This result, which is at odds with what was found in crystalline conductors \cite{marcolongo_ionic_2017,mo_first_2012,adams_structural_2012}, demonstrate that in these materials, the purely alkaline part of the ionic conductivity (i.e. $\sigma_{\textsf{Li}-\textsf{Li}}$) is fundamentally characterized by independent, single-particle diffusion. 
We thus deduce that these materials are characterized by: (i) a sluggish metal-anion framework that constitutes the backbone of the amorphous phase, and (ii) alkali cations that propagate through the interstitial space between metal-anion clusters via a standard hopping process that does not drag neighboring anions with it. Such diffusion mechanism is reminiscent of Li transport in oxide-based superionic conductors, where Li ions migration is promoted when occurring  within a framework of corner-sharing polyhedra \cite{jun_lithium_2022}. We also note that point (i) deviates from what has been observed in Li$-$P$-$S amorphous conductors, where the polyanion building blocks [(PS$_4]^{3-}$, [P$_2$S$_6]^{4-}$ and [P$_2$S$_7]^{4-}$) are found to be immobile \cite{lee_weak_2023,jun_nonexistence_2024}. 

We evaluated the transference number of amorphous oxyhalides using different approximations: the Nernst-Einstein (NE) and the cluster Nernst-Einstein (cNE) approximations, and compared the results to those obtained with the exact Einstein expression of Eq. (\ref{eq::Einstein_tA}). This comparison allows us to disentangle the contributions of different-particles correlations and to progressively include them to evaluate their effects. Under the cluster Nernst–Einstein (cNE) assumption, metal-anion correlations, which are omitted in the simple NE expression, are included. These inter-particle correlations carry negative signs and are large in magnitude; thus, they reduce the total ionic conductivity, yielding higher alkali transference numbers. Within the cNE approach, correlation between the ions constituting polyhedral clusters is included, but the correlation between alkali and polyanions is still neglected; its inclusion through the full Einstein form reveals that such correlation is also sizable, and has the effect of decreasing the transference numbers. Taken together, the results where only metal-anion contributions are retained (cNE approximation) indicate that such ionic correlation plays an important role for optimizing alkaline transport properties. In particular, promoting a more correlated metal--anion dynamics could be a strategy for increasing the alkali transference number: the metal-anion ($M$--$X/\textsf{O}$) contribution to the ionic conductivity is proportional to  $L_{M-X/\textsf{O}} = Z_{M} \,Z_{X/\textsf{O}} \, \mathcal{D}_{M-X/\textsf{O}}$, where $Z$ is the valence charge and $\mathcal{D}_{M-X/\textsf{O}}$ is the slope of the displacement-displacement correlation function 
%$\mathcal{D}_{M-X/\textsf{O}}=\lim_{t\rightarrow\infty}\frac{d}{dt}\langle \Delta r_M\cdot \Delta r_{X/\textsf{O}} \rangle$ 
[see Eq. (\ref{eq::L_terms})]. 
Maximizing the alkali transference number requires $L_{M-X/\textsf{O}}$ to be as negative as possible [see Eq. (\ref{eq::Einstein_tA})]. Since $Z_{X/\textsf{O}}<0$, reducing $L_{M-X/\textsf{O}}$ can be achieved by increasing $Z_M>0$ and/or by enhancing $\mathcal{D}_{M-X/\textsf{O}}>0$. Aliovalent substitution with larger-radius and/or higher-valent metal cations can be a possible way to achieve this route: higher $Z_M$ directly amplifies the charge contribution to $L_{M-X/\textsf{O}}$, while larger ionic radii promote higher coordination numbers, increasing $\mathcal{D}_{M-X/\textsf{O}}$ and thereby facilitating more efficient alkali transport. Recent experimental findings support this hypothesis, reporting transference numbers exceeding 0.89 for $M=$ Ta, Nb, Zr and Hf \cite{yue_universal_2025}.

In summary, our study clarified the structural motifs and ion-transport behavior of amorphous oxyhalide electrolytes, which remains remarkably consistent despite differences in underlying chemistry. The possibility of synthesizing these materials without mechanochemical approaches \cite{dai_inorganic_2023,zhang_universal_2024} also highlights their potential for large-scale production \cite{wang_oxychloride_nodate}. Moreover, the high room-temperature ionic conductivity achieved not only for Li, but also for Na and K cations, indicates that the amorphous oxychloride space can be a fertile source of solid state conductors for beyond-Li technologies. Indeed, upon compositional optimization -- i.e. tuning the oxygen content or employing aliovalent substitution -- it has been shown that it is possible to achieve even higher values of ionic conductivities \cite{Wei2025,ma_coupled_2025}, thereby enabling the development of higher-performance solid electrolytes for all-solid-state batteries.

\section{experimental section}
\textit{Ab initio} calculations have been performed using the Vienna Ab initio Software Package (VASP) \cite{kresse_efficient_1996, kresse_ultrasoft_1999,kresse_ab_1993-1}, where the electron-ion interaction was modeled using projector augmented-wave pseudopotentials \cite{blochl_projector_1994}. The generalized-gradient approximation with the Perdew-Burke-Ernzerhof parametrization was used as parameterization for the exchange-correlation functional \cite{perdew_generalized_1996}. The electronic wavefunction have been expanded with a kinetic-energy cutoff of 500 eV, and $\Gamma$-point sampling of reciprocal space has been adopted. Other input flags follow standard settings as recommended in Pymatgen \cite{ong_python_2013, jain_high-throughput_2011}. Born-Oppenheimer (BO) molecular dynamics (MD) has been used with integration time-step of 2 fs without spin-polarization, using the Nosé–Hoover constant-volume thermostat (NVT) \cite{nose_unified_1984,hoover_canonical_1985} for the melt-and-quench protocol, and the Langevin constant-pressure barostat (NPT) \cite{allen_computer_2017} for generating the fine-tuning dataset. All calculations have been performed using the active-learning methodology implemented in VASP \cite{jinnouchi_--fly_2019,jinnouchi_descriptors_2020,jinnouchi_phase_2019}, where a MLIP is constructed on-the-fly during the MD simulation. Within this computational scheme, a Bayesian inference-based approach is used at each MD step to decide whether to use the constructed MLIP, or to evaluate the ionic forces through electronic self-consistent minimization, which occurs when the Bayesian error exceeds a predefined threshold that is updated along the MD run. This computational methodology is well-suited to efficiently explore the complex BO potential energy surface of amorphous structures (which require an extensive sampling of different atomic configurations), as it allows a factor $\sim$10 speed-up -- thanks to the ML prediction -- but retains an \textit{ab initio} accuracy because only first-principles energies, forces and stresses are used for fine-tuning the MLIP used for MD production runs. For this latter, we used the Crystal Hamiltonian Graph neural Network potential (CHGNet) \cite{deng_chgnet_2023}, that is naturally pretrained on relaxation trajectories from the Materials Project database \cite{jain_commentary_2013}, but which we specifically fine-tuned to model amorphous oxychlorides. 

We fine-tuned the CHGNet model separately for each material. For every system, the fine-tuning dataset consists of \emph{ab initio}-calculated energies, forces, and stresses obtained from on-the-fly MD trajectories of 84-atoms amorphous structures collected at four different temperatures (300, 500, 700, and 1000 K), each of them sampled for 180 ps. To maximize the structural diversity of the training dataset, we also included for every $AMX_{2.5}$O$_{0.75}$ material data from crystalline metal-oxides ($M_2$O$_3$: \texttt{mp-7048}, \texttt{mp-886}, \texttt{mp-22323}), alkaline oxide ($A_2$O: \texttt{mp-1960}, \texttt{mp-2352}, \texttt{mp-971}) and alkaline-metal-halide ($AMX_4$, when available on Materials Project: mp-22983 (LiAlCl$_4$), \texttt{mp-23363} (NaAlCl$_4$), \texttt{mp-27755} (KAlCl$_4$), \texttt{mp-28341} (LiGaCl$_4$)) obtained from a gradual temperature-ramping from 200 to 1100 K in 50 ps. A total of $\sim12\,000$ atomic configurations were collected for each material's dataset. Of these, 95\% have been used for training the ML models and 5\% for the validation set. All other training hyperparameters were fixed to CHGNet default values. The MLIP errors are reported in Table~\ref{tab::ML_errors}. MD production runs have been run using the atomistic simulation environment (ASE) \cite{hjorth_larsen_atomic_2017} and statistical observables concerning diffusion properties -- like the mean square displacement and the mean collective displacement --, together with the associated errors,
have been evaluated using the methodology discussed in He et al. \cite{he_statistical_2018}:
$\big\langle f(t,t=0)\big\rangle=N_\tau^{-1}\sum_{\tau=0}^{T_{\textsf{tot}}-t}f(t+\tau,\tau)$.

To obtain the amorphous model structures, we either started from crystalline $AMX_4$ mp-22983 (LiAlCl$_4$), \texttt{mp-23363} (NaAlCl$_4$), \texttt{mp-27755} (KAlCl$_4$), \texttt{mp-28341} (LiGaCl$_4$), or performed isovalent substitution and further relaxed the resulting structure, and carried out a melt-and-quench protocol as described in Section II, where ramping, equilibration before doping and equilibration after doping lasted for 2, 10 and 20 ps, respectively.

The radial distribution function $g_{A-B}(t)$ and the number of particles $N_{A-B}(t)$ displayed in Figs. \ref{fig::amorphous_structure}(a--c), \ref{fig::residence_time_analysis}(d--f) \ref{fig::residence_time_analysis}(d,e) and \ref{fig::RDF_metals}(a,b) are calculated using the following definitions \cite{hansen90a}: 
\begin{gather}
    g_{A-B}(r)=\frac{1}{\rho_B N_A}\bigg\langle\sum_{i=1}^{N_A}\sum_{j=1}^{N_B}\delta\big(r-|\boldsymbol{r}_i-\bm{r}_j|\big)\bigg\rangle\\ N_{A-B}(R)=\rho_B\,\int_0^{R}\,dr\,4\pi r^2\,\,g_{A-B}(r).
\end{gather}
During the preparation of this work, the authors used ChatGPT to edit the final draft of the manuscript. After using this tool, the authors reviewed and edited the content as needed and they take full responsibility for the content of the publication.

\section{acknowledgements}
L. B. acknowledges financial support by Umicore, Contract Number 34586 to the Regents of the University of California, Berkeley. This work was supported by the computational resources provided by the Kestrel and Swift clusters from the National Renewable Energy Laboratory (NREL).

\section{Conflict of Interest}
The authors declare no conflict of interest.

\section{Data Availability Statement}
The data that support the findings of this study are available from the corresponding author upon reasonable request.

\section{Keywords}
Superionic conductors, machine learning interatomic potentials, amorphous materials, diffusion dynamics, ionic correlations.

\bibliography{references, references-2}

\clearpage
\beginsupplement     % reset and change numbering to S1, S2,...

% include the content of SI.tex (SI.tex must NOT contain its own preamble)
\onecolumngrid
\begin{center}
  {\large Supplementary Information for}\\[4pt]
  {\large\bfseries\MainTitle}\\[12pt]
  Luca Binci,$^{1,2}$ KyuJung Jun,$^{1,2}$ Bowen Deng$^{1,2}$ and Gerbrand Ceder$^{1,2}$\\[5pt]
  $^1${\itshape Department of Materials Science and Engineering, 
\\ University of California Berkeley, Berkeley, California, USA}\\
  $^2${\itshape Materials Sciences Division, Lawrence Berkeley National Laboratory, Berkeley, California, USA}\\[2pt]
  (Dated: \today)
\end{center}

\clearpage    

\begin{figure}[]
\centering
\includegraphics[width=0.62\textwidth]{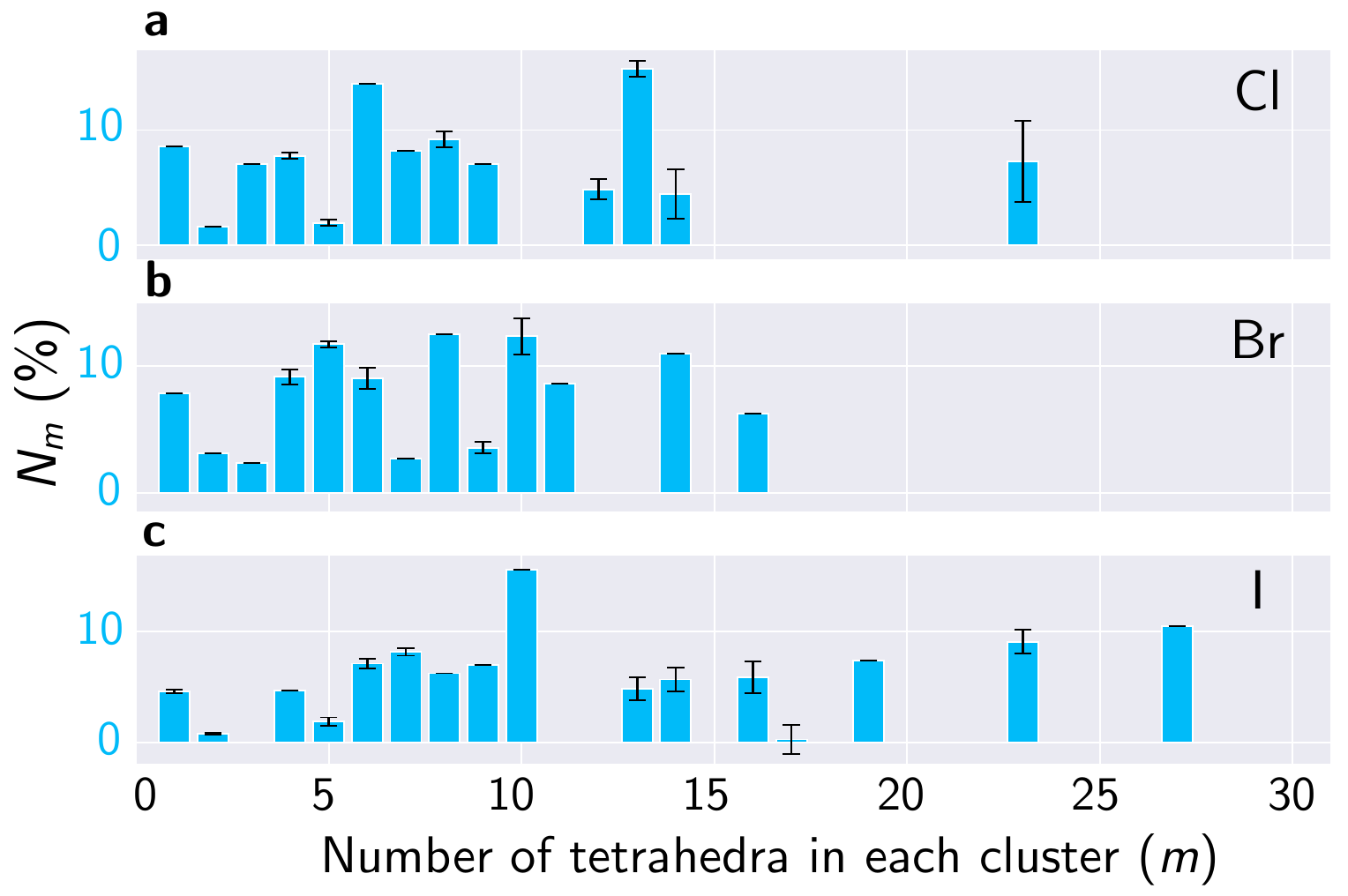}
\caption{\justifying Same as Fig. \ref{fig::amorphous_structure}\textbf{(a--c)}: Barplot of the average percentage of metal–anion tetrahedral units ($N_m$) versus the number of tetrahedral units ($m$) per cluster. Results are shown for LiAl$X_{2.5}$O$_{0.75}$ ($X=$ Cl {[a]}, Br {([b]}, I {[c]}). Error bars represent the standard deviation $\Delta_m$ (defined in the main text)..\label{fig::cluster_halogens}}
\end{figure} 

\begin{figure}[]
\centering
\includegraphics[width=0.55\textwidth]{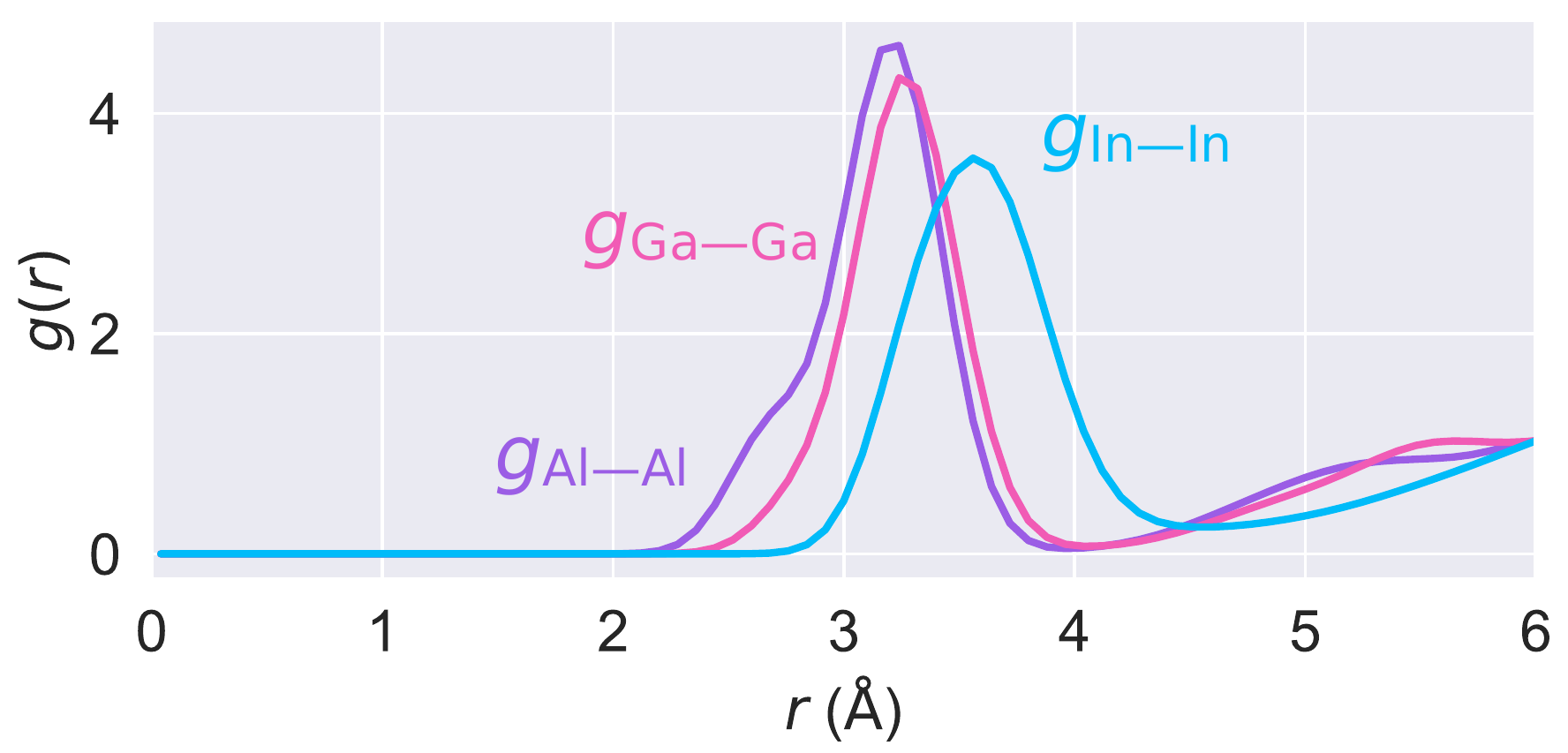}
\caption{ Radial distribution function $g_{M-M}(r)$ between metal cations $M=$ Al (purple), Ga (pink), In (light blue).\label{fig::g_r_metal-metal}}
\end{figure}

\begin{figure}[]
\centering
\includegraphics[width=0.85\textwidth]{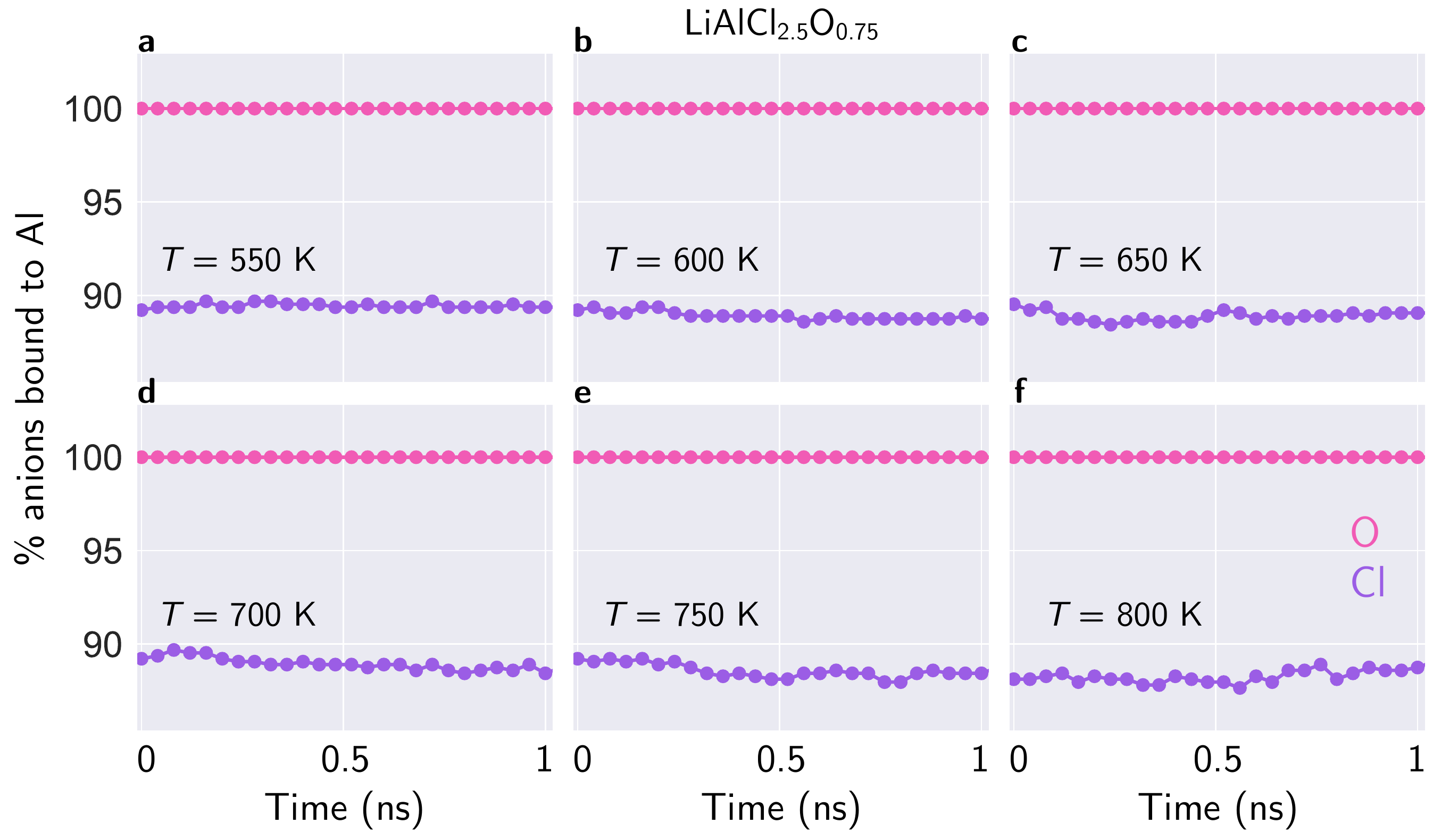}
\caption{\justifying Fraction of oxygen (pink) and chlorine (purple) bound to Al; i.e. found within a sphere of radius of 1.3 $\textsf{\AA}$ (O) and 2.7 $\textsf{\AA}$ (Cl) from aluminum cations (as taken from the $g(r)$ in Fig. \ref{fig::residence_time_analysis}d) for different temperatures (550 \textbf{[a]}, 600 \textbf{[b]}, 650 \textbf{[c]}, 700 \textbf{[d]}, 750 \textbf{[e]}, 800 \textbf{[f]}) in LiAlCl$_{2.5}$O$_{0.75}$.\label{fig::bound_anions}}
\end{figure} 

\begin{figure}[]
\centering
\includegraphics[width=0.9\textwidth]{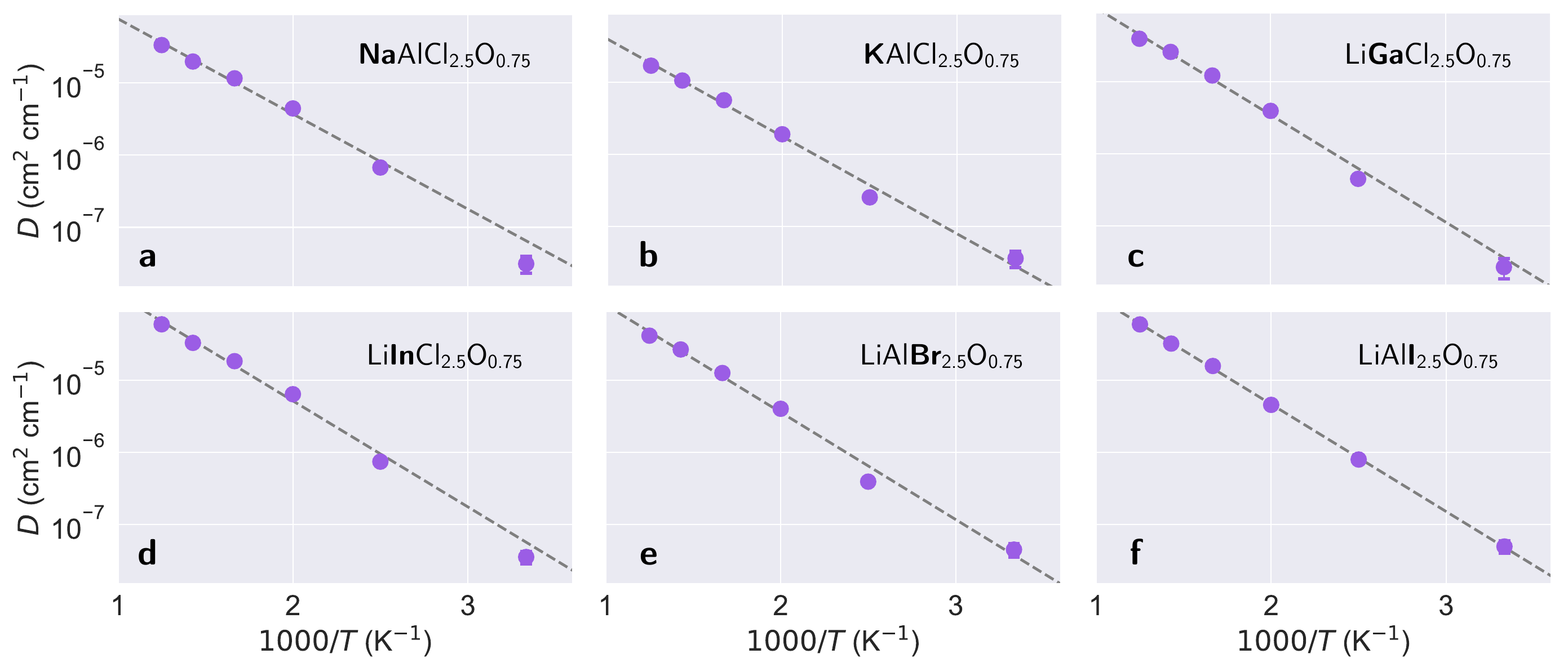}
\caption{\justifying Arrhenius plot of $D$ as a function of the inverse temperature for $AMX_{2.5}$O$_{0.75}$ ($A=$ Na \textbf{[a]}, K \textbf{[b]}; $M=$ Ga \textbf{[c]}, In \textbf{[d]}; $X=$ Br \textbf{[e]}, I \textbf{[f]}); the dashed line is the linear fit used to extract the activation energy. Results are shown for all the oxyhalides except for LiAlCl$_{2.5}$O$_{0.75}$ (which is reported in the main text).\label{fig::all_arrhenius}}
\end{figure}

\begin{figure}[]
\centering
\includegraphics[width=0.95\textwidth]{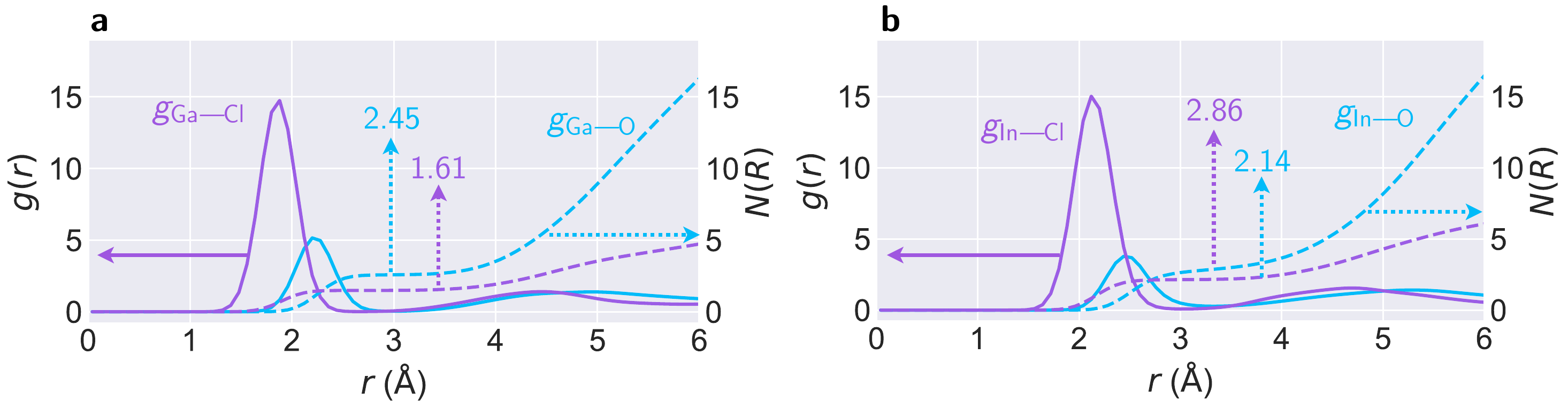}
\caption{\justifying Radial distribution functions $g_{B-B}(r)$ (continuous lines) and average particle number $N_{M-B}(r)$ (dashed lines) between a metal $M$ and anion $B$ of Li$M$Cl$_{2.5}$O$_{0.75}$ with $M=$ Ga in panel \textbf{(a)} and $M=$ In in panel \textbf{(b)}.\label{fig::RDF_metals}} 
\end{figure} 

\begin{figure}[]
\centering
\includegraphics[width=0.77\textwidth]{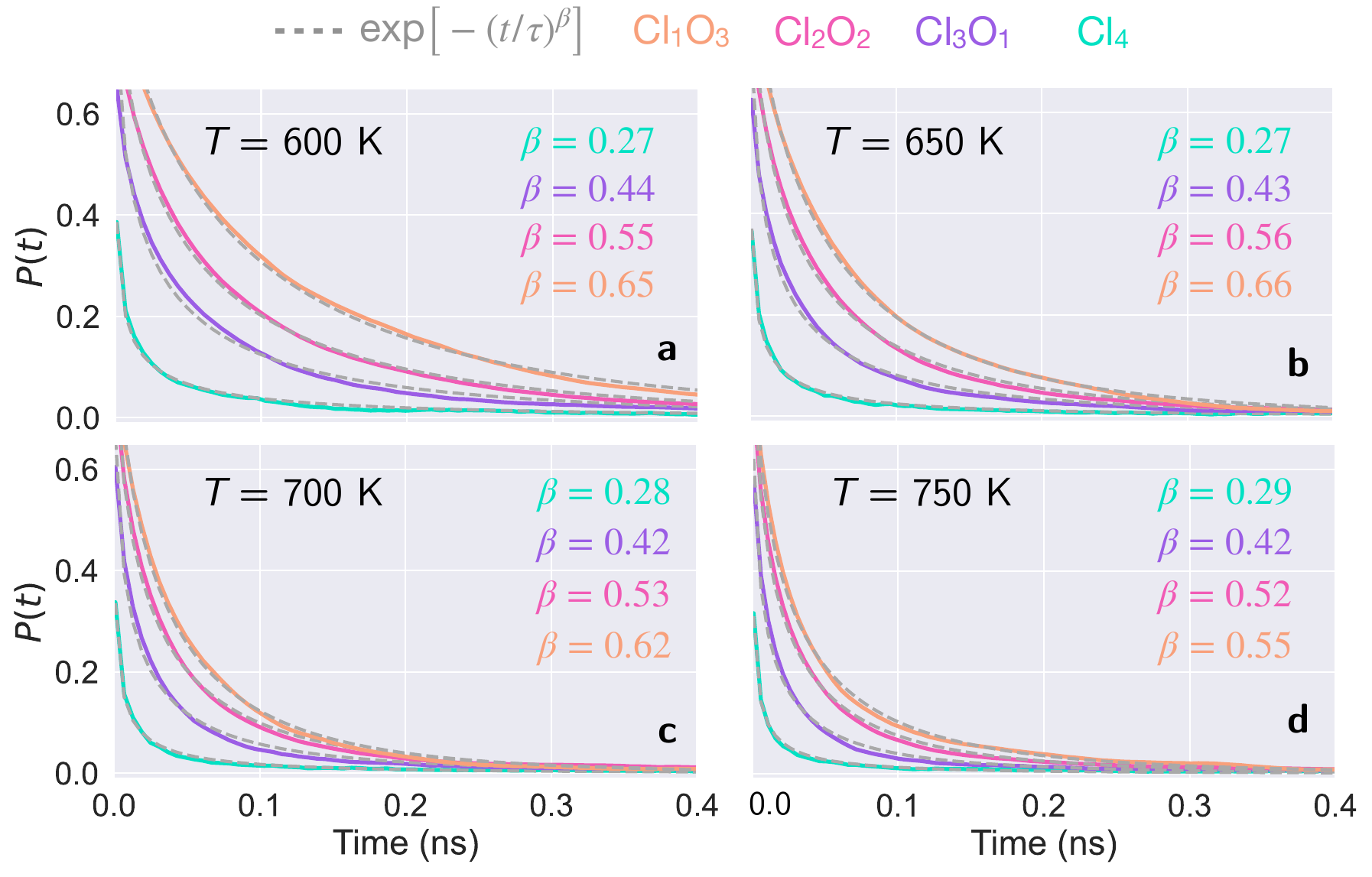}
\caption{\justifying Calculated $P(t)$ (continuous lines) and associated exponential fit (dashed lines) of LiAlCl$_{2.5}$O$_{0.75}$ at different temperatures: 600 K \textbf{(a)}, 650 K \textbf{(b)}, 700 K \textbf{(c)}, 750 K \textbf{(d)}. The inset show the associated values of $\beta$ extracted from the stretched exponential fit $\exp\,[-(t/\tau)^\beta]$.\label{fig::beta}} 
\end{figure} 

\begin{figure}[]
\centering
\includegraphics[width=0.95\textwidth]{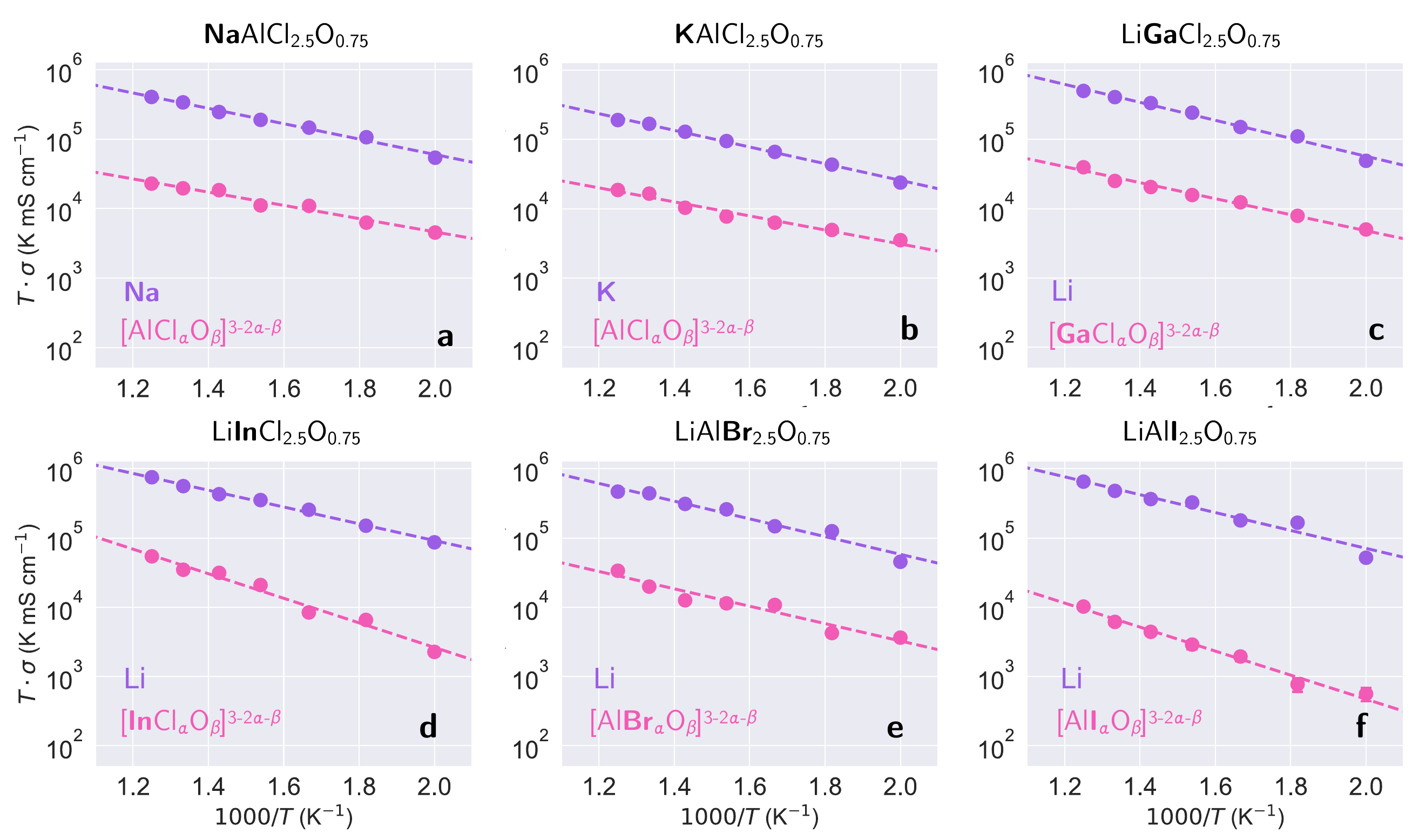}
\caption{\justifying Conductivities (multiplied by their corresponding temperatures) as a function of the inverse temperature of alkali (Li, Na \textbf{[a]}, K \textbf{[b]} -- purple points), calculated with the NE approximation, and of [$M$$X_\alpha$O$_\beta$]$^{3-\alpha-2\beta}$ clusters ($M=$ Ga \textbf{[c]}, In \textbf{[d]}; $X=$ Br \textbf{[e]}, I \textbf{[f]} -- pink points), evaluated within the cNE approximation. Dashed lines are linear fit. Results are shown for all the oxyhalides except for LiAlCl$_{2.5}$O$_{0.75}$ (which is reported in the main text).\label{fig::cNE_series}} 
\end{figure} 

\begin{figure}[]
\centering
\includegraphics[width=0.95\textwidth]{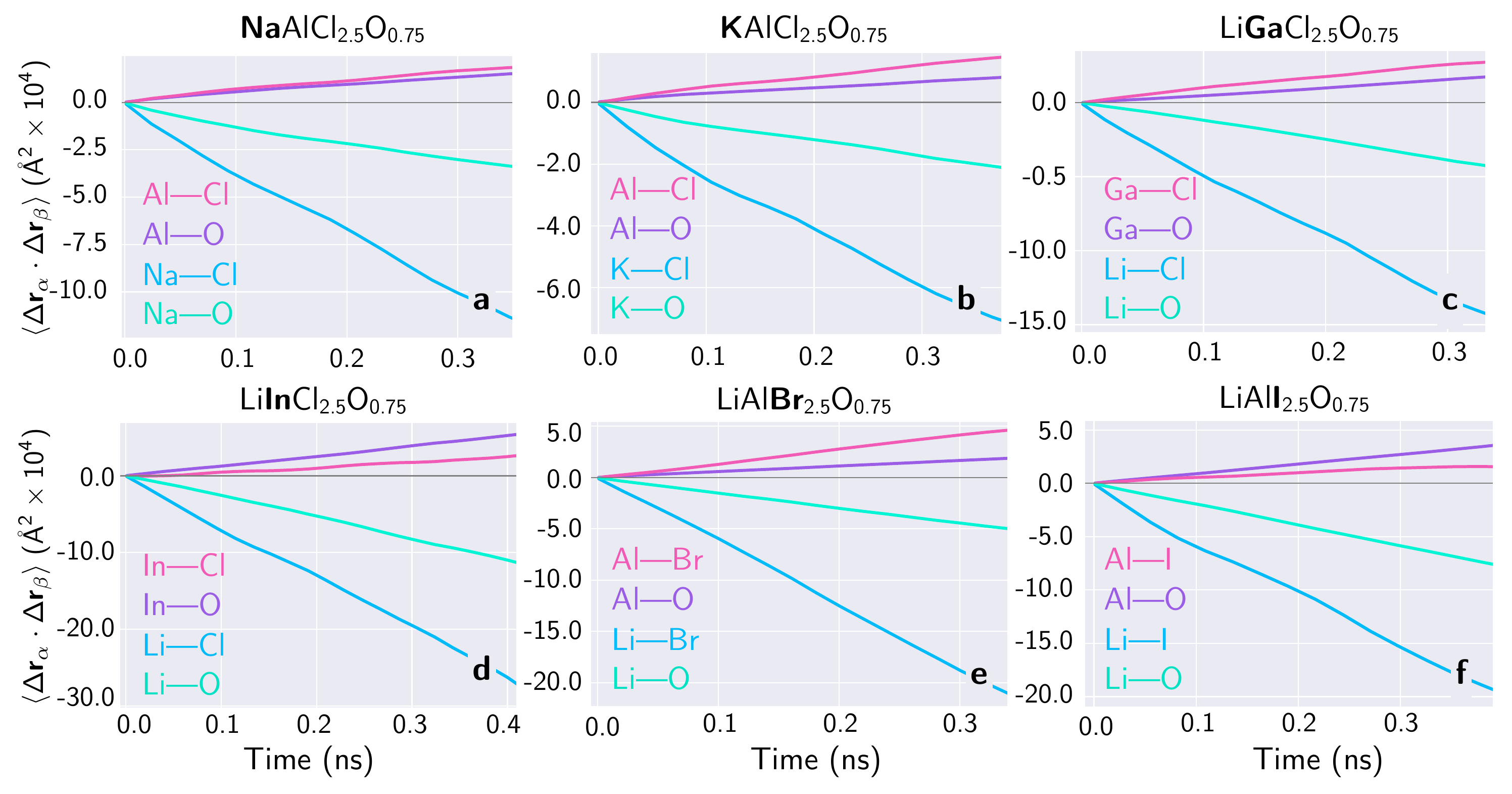}
\caption{\justifying Displacement-displacement correlation function for $AMX_{2.5}$O$_{0.75}$ ($A=$ Na \textbf{[a]}, K \textbf{[b]}; $M=$ Ga \textbf{[c]}, In \textbf{[d]}; $X=$ Br \textbf{[e]}, I \textbf{[f]}) between alkali and anions (light blue and turquoise) and metals and anions (purple and pink).} 
\end{figure}

\begin{figure}[]
\centering
\includegraphics[width=0.45\textwidth]{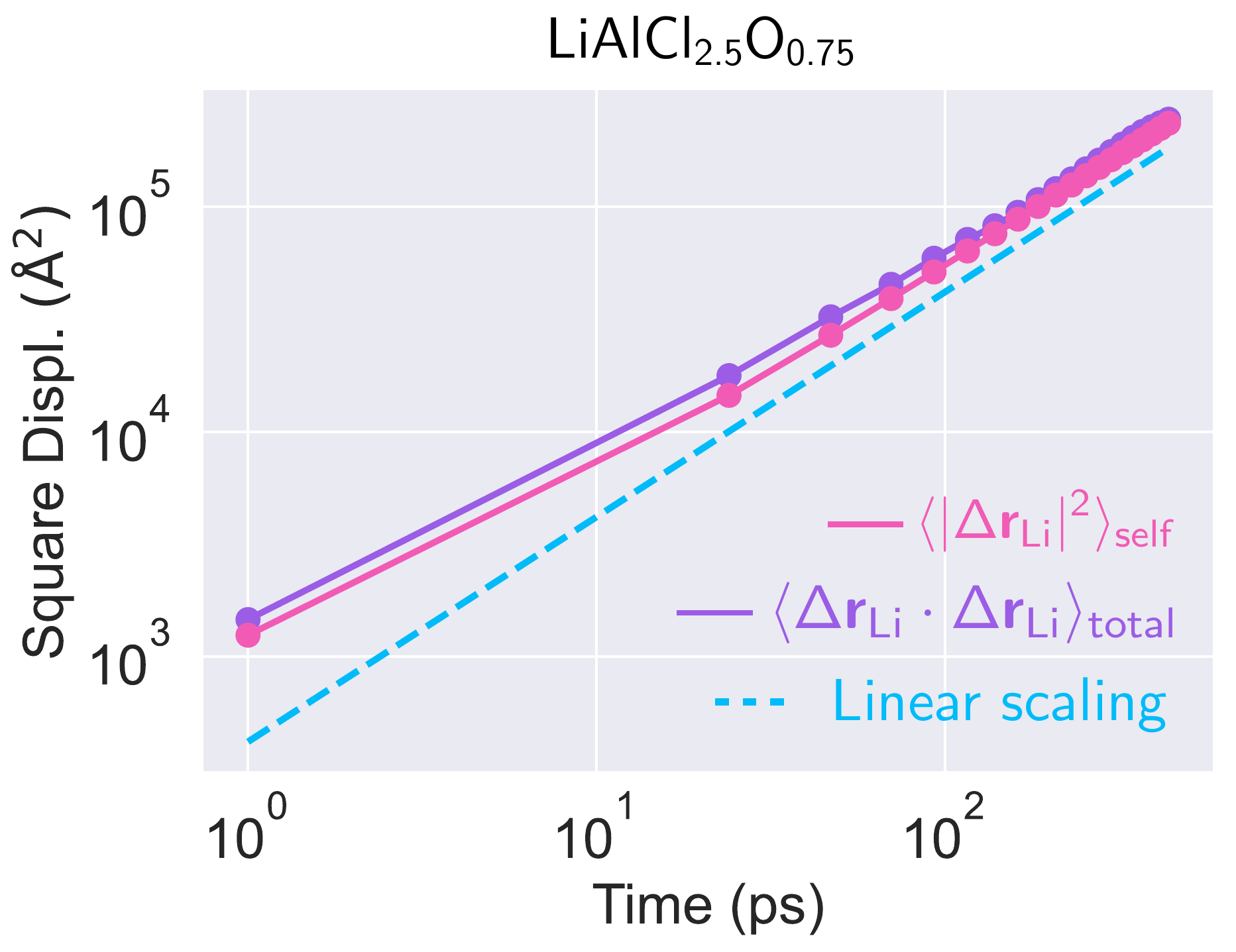}
\caption{\justifying Same data as shown in Fig. \ref{fig::onsager}(b) for the total and self part of the lithium collective mean displacement, but in log-log scale. The dashed blue line indicates the ideal linear scaling.\label{fig::log-log}} 
\end{figure}

\begin{figure}[]
\centering
\includegraphics[width=0.95\textwidth]{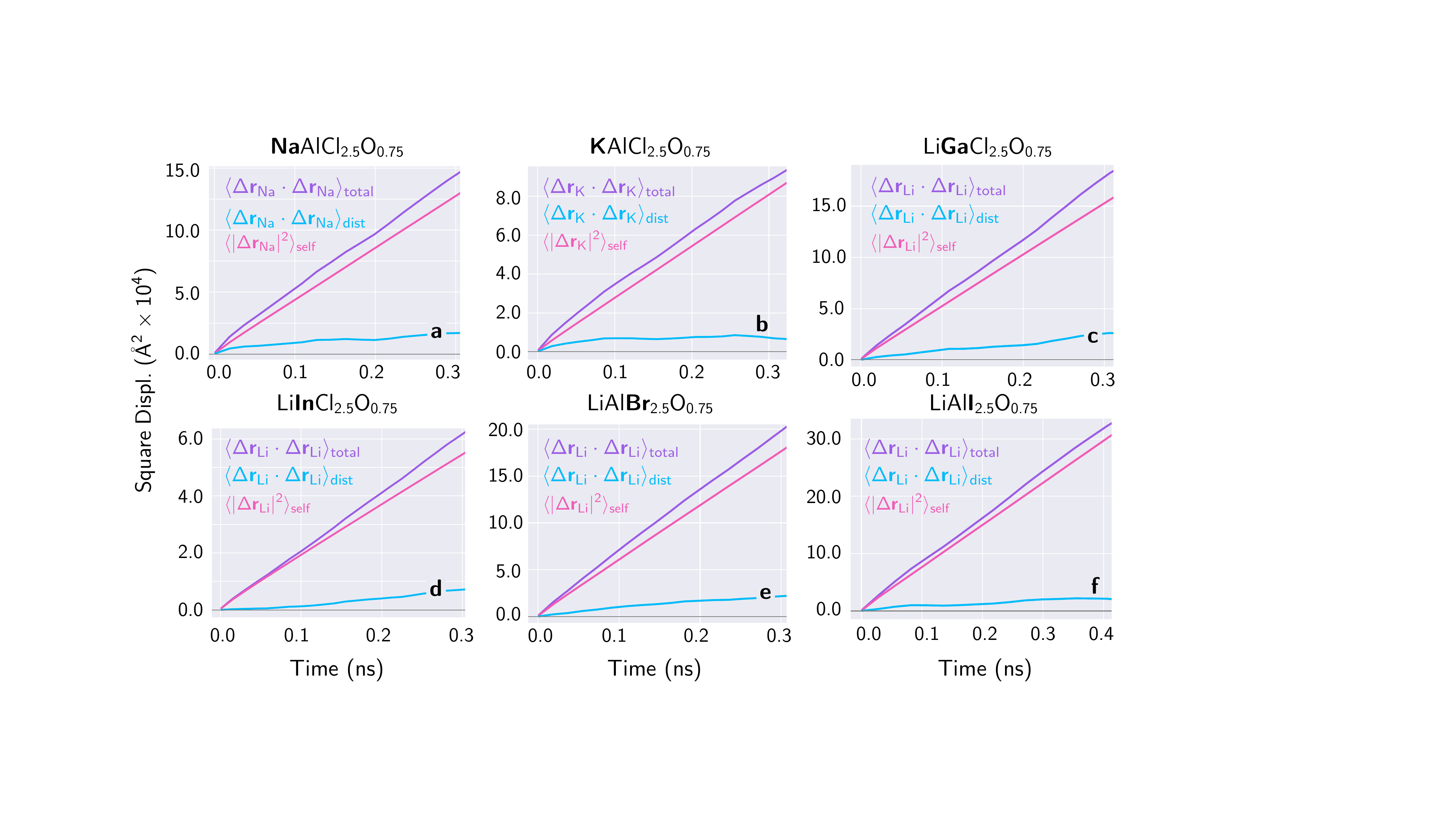}
\caption{\justifying Alkali-alkali ($A$$-$$A$) displacement-displacement correlation function (DDCF)(purple line), mean square displacement multiplied by $N^A$ (pink line), and distinct part of $A$$-$$A$ DDCF (light blue line) in $AMX_{2.5}$O$_{0.75}$ ($A=$ Na \textbf{[a]}, K \textbf{[b]}; $M=$ Ga \textbf{[c]}, In \textbf{[d]}; $X=$ Br \textbf{[e]}, I \textbf{[f]}); the distinct part represents the off-diagonal elements of the $A$$-$$A$ DDCF. \label{fig::for_haven}} 
\end{figure}

\begin{table}[]
    \centering
\renewcommand{\arraystretch}{1.}
    \setlength\tabcolsep{0.38in}
\caption{\justifying Machine learning errors on energies, forces and stresses for the investigated materials.}\label{tab::ML_errors}
\begin{tabular}{cccc}
        \toprule
        \multirow{2}{*}{Material}&\multicolumn{3}{c}{Errors}\\
        &Energies (meV/atom)&Forces (meV/\AA)&Stresses (GPa)\\
        \midrule
        LiAlCl$_{2.5}$O$_{0.75}$&2&46&0.037\\
        NaAlCl$_{2.5}$O$_{0.75}$&4&57&0.075\\
        KAlCl$_{2.5}$O$_{0.75}$&4&52&0.061\\
        LiGaCl$_{2.5}$O$_{0.75}$&3&49&0.043\\
        LiInCl$_{2.5}$O$_{0.75}$&2&48&0.033\\
        LiAlBr$_{2.5}$O$_{0.75}$&2&47&0.034\\
        LiAlI$_{2.5}$O$_{0.75}$&3&55&0.034\\
        \bottomrule
    \end{tabular}
\end{table}

\begin{table}[]
    \centering
\renewcommand{\arraystretch}{0.7}
    \setlength\tabcolsep{0.3in}
\caption{\justifying Calculated temperature-dependent conductivities $\sigma$  (in mS cm$^{-1}$) within the Nernst-Einstein approximation for all the materials investigated in this work.}\label{sigma_NE}
\begin{tabular}{cccccccc}
        \toprule
        Ion & $\sigma_{500}$ & $\sigma_{550}$ & $\sigma_{600}$ & $\sigma_{650}$ & $\sigma_{700}$ & $\sigma_{750}$ & $\sigma_{800}$\\
        \midrule
        \multicolumn{8}{c}{LiAlCl$_{2.5}$O$_{0.75}$}\\
        Li & 83 & 193 & 242 & 357 & 441 & 526 & 675\\
        Al & 71 & 145 & 169 & 215 & 334 & 352 & 405\\
        Cl & 29 & 64  & 70  & 109 & 139 & 189 & 209\\
        O  & 18 & 42 & 40 & 59 & 93 & 85 & 107\\
        \midrule
        \multicolumn{8}{c}{NaAlCl$_{2.5}$O$_{0.75}$}\\
        Na & 109 & 191 & 244 & 294 & 354 & 452 & 511\\
        Al & 85 & 105 & 167 & 165 & 237 & 251 & 269 \\
        Cl & 35 & 57 & 71 & 85 & 113 & 138 & 163\\
        O & 22 & 25 & 46 & 40 & 65 & 65 & 66 \\
        \midrule
        \multicolumn{8}{c}{KAlCl$_{2.5}$O$_{0.75}$}\\
        K & 49 & 77 & 109 & 144 & 186 & 226 & 237\\
        Al & 64 & 83 & 102 & 113 & 142 & 202 & 213 \\
        Cl & 25 & 39 & 43 & 59 & 65 & 94 & 108\\
        O & 18 & 21 & 28 & 30 & 37 & 54 & 54 \\
        \midrule 
        \multicolumn{8}{c}{LiGaCl$_{2.5}$O$_{0.75}$}\\
        Li & 99 & 203 & 256 & 372 & 473 & 545 & 635\\
        Ga & 95 & 135 & 198 & 235 & 280 & 321 & 444 \\
        Cl & 37 & 63 & 88 & 117 & 141 & 169 & 219\\
        O & 25 & 32 & 53 & 59 & 73 & 82 & 120 \\
        \midrule 
        \multicolumn{8}{c}{LiInCl$_{2.5}$O$_{0.75}$}\\
        Li & 170 & 273 & 425 & 542 & 624 & 755 & 945\\
        In & 45 & 114 & 127 & 294 & 403 & 420 & 625 \\
        Cl & 60 & 120 & 178 & 265 & 381 & 475 & 645\\
        O & 9 & 28 & 22 & 64 & 69 & 90 & 71 \\
        \midrule 
        \multicolumn{8}{c}{LiAlBr$_{2.5}$O$_{0.75}$}\\
        Li & 77 & 228 & 233 & 402 & 393 & 587 & 663\\
        Al & 57 & 78 & 150 & 173 & 293 & 251 & 453 \\
        Br & 24 & 54 & 59 & 113 & 124 & 159 & 200\\
        O & 14 & 21 & 43 & 45 & 88 & 95 & 131 \\
        \midrule 
        \multicolumn{8}{c}{LiAlI$_{2.5}$O$_{0.75}$}\\
        Li & 89 & 305 & 283 & 496 & 496 & 650 & 801\\
        Al & 17 & 16 & 48 & 43 & 113 & 122 & 116 \\
        I & 11 & 20 & 32 & 52 & 75 & 90 & 125\\
        O & 5 & 5 & 13 & 10 & 33 & 21 & 32 \\
        \bottomrule
    \end{tabular}
\end{table}

\begin{table}[]
    \centering
\renewcommand{\arraystretch}{1.}
    \setlength\tabcolsep{0.25in}
\caption{Some representative calculated coefficients $L_{\alpha\beta}$ (in \AA$^2$/ps) as in Eq. (\ref{eq::Einstein}) at $T=750$~K .}\label{table::einstein_coeff}
\begin{tabular}{cccccccc}
        \toprule
        \multicolumn{8}{c}{$L_{\alpha\beta}$ }\\
        $\alpha$--$\beta$ & LACO & NACO & KACO & LGCO & LICO & LABO & LAIO\\
        \midrule
        $A$--$A$& 709& 403& 228& 558& 940& 757& 558\\
        $A$--$M$& -638& -366& -212& -495& -917& -590& -520\\
        $A$--$X$& 561& 288& 158& 396& 602& 650& 394\\
        $A$--O&   252& 179& 103& 257& 463& 263& 244\\
        $M$--$M$& 659& 459& 239& 553& 1195& 567& 531\\
        $M$--$X$& -416& -121& -109& -220& -268& -396& -327\\
        $M$--O& -281&  -255& -117& -318& -652& -274& -248\\
        $X$--$X$& 560&376& 157& 465& 756& 694& 317\\
        $X$--O& 114& 8& 38& 42& 43& 132& 151\\
        O--O& 160& 167& 75& 236& 409& 163& 121\\
        \bottomrule
    \end{tabular}
\end{table}

\end{document}